\documentclass[12pt]{article}
\usepackage{graphicx}
\usepackage{enumitem}
\usepackage{natbib} 
\usepackage{url} 
\usepackage{amsmath} 
\usepackage{amsfonts}
\usepackage{amssymb} 
\usepackage{lineno}
\usepackage{setspace}
\usepackage{lscape} 
\usepackage{float} 
\usepackage{breakcites}
\usepackage{pdfpages}
\usepackage{xcolor}
\usepackage{hyperref}
\usepackage{setspace}
\usepackage{graphicx}
\usepackage{natbib}
\usepackage{url}
\usepackage{amsmath}
\usepackage{amsfonts}
\usepackage{amssymb}
\usepackage{lineno}
\usepackage{setspace}
\usepackage{lscape} 
\usepackage{float} 
\usepackage{breakcites}
\usepackage{pdfpages}
\usepackage{bm}
\usepackage{xcolor}
\usepackage{hyperref}
\usepackage{setspace}
\usepackage{amsmath}
\usepackage{amssymb}
\usepackage{graphicx}
\usepackage{mathtools}
\usepackage{float}
\usepackage{tabularx}
\usepackage{array}
\usepackage{wrapfig}
\usepackage{multirow}
\usepackage{tabu}
\usepackage{booktabs}
\usepackage{pgfplots}
\usepackage{pgfplotstable}
\usepackage{qtree}
\usepackage{xcolor}
\usepackage{listings}
\usepackage{caption}
\usepackage{siunitx}
\usepackage{makecell}
\usepackage{bm}
\usepackage{setspace}
\usepackage{subfig}
\usepackage{titlesec}
\usepackage{url}
\usepackage{colortbl}
\usepackage{dsfont}
\usepackage{amsthm}

\graphicspath{{Figures/}}
\newcommand{\blind}{0}

\usepackage{algorithm,algpseudocode}

\begin{document}

\def\spacingset#1{\renewcommand{\baselinestretch}%
{#1}\small\normalsize} \spacingset{1}

\newtheorem{theorem}{Theorem}
\newtheorem{proposition}{Proposition}

\if0\blind
{
  \title{\bf Flexible Basis Representations for Modeling Large Non-Gaussian Spatial Data}
\author{Remy MacDonald and Benjamin Seiyon Lee\hspace{.2cm}\\
Department of Statistics, George Mason University}
\date{}
\maketitle
} \fi

\if1\blind
{
  \bigskip
  \bigskip
  \bigskip
  \begin{center}
    {\LARGE\bf Adapt-BaSeS: Flexible Basis Representations for Modeling Large Non-Gaussian Spatial Datasets}
\end{center}
  \medskip
} \fi

\begin{abstract}
Nonstationary and non-Gaussian spatial data are common in various fields, including ecology (e.g., counts of animal species), epidemiology (e.g., disease incidence counts in susceptible regions), and environmental science (e.g., remotely-sensed satellite imagery). Due to modern data collection methods, the size of these datasets have grown considerably. Spatial generalized linear mixed models (SGLMMs) are a flexible class of models used to model nonstationary and non-Gaussian datasets. Despite their utility, SGLMMs can be computationally prohibitive for even moderately large datasets (e.g., 5,000 to 100,000 observed locations). To circumvent this issue, past studies have embedded nested radial basis functions into the SGLMM. However, two crucial specifications (knot placement
and bandwidth parameters), which directly affect model performance, are typically fixed prior to model-fitting. We propose a novel approach to model large nonstationary and non-Gaussian spatial datasets using adaptive radial basis functions. Our approach: (1) partitions the spatial domain into subregions; (2) employs reversible-jump Markov chain Monte Carlo (RJMCMC) to infer the number and location of the knots within each partition; and (3) models the latent spatial surface using partition-varying and adaptive basis functions. Through an extensive simulation study, we show that our approach provides more accurate predictions than competing methods while preserving computational efficiency. We demonstrate our approach on two environmental datasets - incidences of plant species and counts of bird species in the United States.
\end{abstract}

\noindent%
\vfill

\newpage
\spacingset{2}

\setlength{\belowdisplayskip}{3pt} \setlength{\belowdisplayshortskip}{3pt}
\setlength{\abovedisplayskip}{3pt} \setlength{\abovedisplayshortskip}{3pt}

\section{Introduction}
\label{intro}

Discrete non-Gaussian spatial datasets (counts, binary responses, extreme values) are prevalent across a number of disciplines, such as ecology \citep{guan2018computationally}, public health \citep{ejigu2020geostatistical}, and atmospheric sciences \citep{sengupta2016predictive,heaton2017nonstationary}. Modeling such datasets can be important for scientific applications, particularly in making predictions at unobserved locations and assessing prediction uncertainty. However, traditional regression models, which assume independent and identically distributed errors, may be inappropriate for these data \citep{schabenberger2017statistical,banerjee2003hierarchical,cressie1993} as they neglect spatial autocorrelation.

Spatial generalized linear mixed models (SGLMMs) \citep{diggle1998model,haran2011gaussian} are a flexible class of spatial models that extend to non-Gaussian observations. Within SGLMMs, the spatial dependence is captured via location-specific random effects that are modeled as a latent Gaussian process (GP). Despite their flexibility, standard implementation of SGLMMs incurs a computational cost that is cubic in the data size, which can be computationally prohibitive for modeling modern spatial datasets. Additionally, the high-dimensional spatial random effects are typically highly correlated; thereby resulting in slow mixing Markov chain Monte Carlo (MCMC) algorithms \citep{haran2003accelerating}.

Computationally-efficient approaches have been developed to reduce the dimensionality of the spatial random effects, speed up large matrix operations, or both. These include low-rank approximations and basis representations \citep{cressie2006spatial, banerjee2008gaussian,finley2009improving}, sparse covariance and precision matrices \citep{furrer2006covariance,datta2016hierarchical,vecchia1988estimation, zilber2021vecchia}, spatial partial differential equations \citep{lindgren2011explicit}, spatial partitioning \citep{lee2023scalable,heaton2017nonstationary}, and more. Two prominent examples include nearest-neighbor Gaussian processes (NNGP) \citep{datta2016hierarchical} and Integrated nested Laplace approximations (INLA) \citep{lindgren2015bayesian}. NNGP approximates the GP using a sparse Cholesky factorization of the specified precision matrix. Sparsity is induced by directed acyclical graphs that connect neighboring locations. While NNGP preserves the flexibility and interpretability of GP models, the SGLMM framework requires the inference of all $n$ spatial random effects. This can be prohibitive for large discrete non-Gaussian spatial datasets. Moreover, the spNNGP \citep{finley2009improving} package currently accommodates binary and Gaussian spatial data only; hence, it cannot directly model count observations \citep{dovers2023fast}. INLA employs stochastic partial differential equations to provide fast and accurate numerical approximations of posterior distributions. INLA provides approximations of the marginal posterior distribution, rather than the joint distribution; hence, it may potentially underestimate the uncertainty in estimation and predictions \citep{ferkingstad2015improving}.

In this manuscript, we focus on basis representation approaches which approximate the latent spatial process using a linear combination of spatial radial basis functions (e.g., bisquare, Wendland basis, and thin-plate-splines) \citep{zammit2017frk, sengupta2013hierarchical, katzfuss2017multi, cressie2008fixed, lee2023scalable}. For a comprehensive review of basis functions models in spatial statistics, please see \cite{cressie2022basis}. Two key components of radial basis functions are the knots (centers of the basis functions) and the associated bandwidth (or smoothing) parameters. The bandwidth defines the ``spread'' of the radial basis function and also tunes the tradeoff between the goodness-of-fit and the roughness of the resulting basis approximation \citep{kato2009model}.  
Poorly specified knots and bandwidths can result in inaccurate representations of the latent spatial surface \citep{sheather1991reliable}. Since specifying these parameters can be challenging, past studies   \citep{cressie2008fixed,katzfuss2011spatio,katzfuss2012bayesian} have typically fixed them prior to model fitting. As a result, the spatial basis functions are constructed without any feedback or influence from the observed data.

To address these challenges, we propose a computationally efficient, yet flexible approach for modeling nonstationary non-Gaussian spatial data. Our method addresses the two limitations (knot placement and bandwidth specification) by allowing the spatial radial basis functions to adapt to the observations. Our method partitions the spatial domain into disjoint subregions and allows the bandwidths to vary across each subregion. For each partition, we employ a reversible jump Markov chain Monte Carlo (RJMCMC) algorithm \citep{green1995reversible} to infer the number and placement of knots. The proposed approach allows for more flexibility than using fixed basis functions and scales well to large datasets.

Previous studies have focused on either adaptive knot selection \citep{biller2000adaptive,denis2010free,katzfuss2013bayesian} or adaptive bandwidth selection \citep{politis2003adaptive,brockmann1993locally}. To the best of our knowledge, our study is the first to consider both adaptive bandwidth and adaptive knot selection simultaneously, especially within spatial models. While adaptive knot selection and bandwidth specification have been applied in splines and kernel regression, their use in spatial statistics remains limited. Additionally, our method differs from existing approaches by allowing partition-specific basis functions instead of assuming globally defined bases.

The outline of the remainder of the paper is as follows. In Section~\ref{subsec:sglmm}, we introduce SGLMMs and basis-expansion SGLMMs and discuss important modeling and computational challenges. In Section~\ref{sec:methodology}, we propose our approach (Adapt-BaSeS) and provide implementation details. We demonstrate our approach via a simulation study in Section~\ref{subsec:SimulationStudy} and real-world applications in Section~\ref{subsec:Applications}. Concluding remarks and directions for future research are provided in Section~\ref{subsec:discussion}.

\section{Spatial Generalized Linear Mixed Models}
\label{subsec:sglmm}
SGLMMs \citep{diggle1998model} are a class of flexible models for modeling spatially-dependent non-Gaussian data. These models are a special case of generalized linear mixed models, where the random effects exhibit spatial correlation. Conditioned on the random effects, the observations are assumed to be independent and follow a location-specific probability distribution. SGLMMs have been used extensively in the literature to model non-Gaussian spatially-correlated data \citep{hughes2013dimension,zilber2021vecchia,zhang2002estimation}.

In the Bayesian framework, SGLMMs are part of the broader class of Bayesian hierarchical spatial models (HSM) \citep{wikle1998hierarchical}. HSMs are comprised of three layers or models: data, process, and parameters. The data model layer contains the probability distribution of the observed data (e.g., Gaussian, Bernoulli, multinomial) conditioned on the latent spatial Gaussian random fields and data model parameters. The process model layer includes conditional distributions for the Gaussian spatial process(es) conditioned on the process model parameters or other underlying processes. Note that the process models can be as simple (single latent process) or robust (multiple interconnected spatial random processes) as needed. Bayesian hierarchical models are applicable for modeling complex spatial processes where: (1) the spatial dependence is driven by a complex latent spatial random process; and (2) the practitioner has prior knowledge of the unknowns (e.g., model parameters and spatial processes).

Let $\{z(\boldsymbol s):\boldsymbol s\in\mathcal D\}$ denote the non-Gaussian observations on the spatial domain $\mathcal D\subset\mathbb R^d$, $d\in\mathbb N$.
At $n$ locations, we have observations $\boldsymbol z=(z(\boldsymbol s_1),\ldots, z(\boldsymbol s_n))^\top$, where $z(\cdot)\sim F(\cdot)$ for some distribution $F$. The conditional mean is modeled as 
$g(\mathbb E[z(\boldsymbol s_i)]):=\eta(\boldsymbol s_i)$ for $i=1,\ldots, n$, where $g(\cdot)$ is a link function and $\eta(\cdot)$ is the linear predictor. For location $\boldsymbol s_i$, the linear predictor is defined as,
$$\eta(\boldsymbol s_i):=\boldsymbol x(\boldsymbol s_i)^\top\boldsymbol\beta+w(\boldsymbol s_i),$$

\noindent where $\boldsymbol x(\boldsymbol s_i)$ is a vector of covariates with regression coefficients $\boldsymbol\beta$, and $w(\boldsymbol s_i)$ represents the spatially-correlated random effect, often modeled as a zero-mean GP $w(\cdot)\sim\mathcal{GP}(\mathbf{0},K)$, where $K$ is a covariance function with marginal variance $\sigma^2$ and a correlation function $C$, i.e., $K(\boldsymbol s_1, \boldsymbol s_2)=\sigma^2 C(\boldsymbol s_1,\boldsymbol s_2),\boldsymbol s_1,\boldsymbol s_2\in\mathcal D$. The correlation function $C:(\mathcal D\times\mathcal D)\rightarrow[-1,1]$ is assumed to be known up to some parameters $\boldsymbol\theta$. A commonly used class of covariance functions, which assumes stationarity and isotropy, is the Mat\'ern class \citep{williams2006gaussian}, defined as,
$$\mathcal M_{\nu}(h)=\frac{2^{1-\nu}}{\Gamma(\nu)}\left(\sqrt{2\nu}\frac{d}{\rho}\right)^{\nu}\mathcal K_{\nu}\left(\sqrt{2\nu}\frac{d}{\rho}\right),\quad\nu,\rho>0,$$
\noindent where  $d= ||\boldsymbol s_i-\boldsymbol s_j||$ denotes the Euclidean distance between pairs of locations, $\rho$ is the spatial range parameter, $\nu$ is the smoothness parameter, $\Gamma(\cdot)$ is the gamma function, and $\mathcal K_{\nu}(\cdot)$ is the modified Bessel function of the second kind of order $\nu > 0$.

\sloppy

For a finite vector of locations $\mathcal S=(\boldsymbol s_1,\ldots,\boldsymbol s_n)$, the spatial random effects $\boldsymbol w=(w(\boldsymbol s_1),\ldots,w(\boldsymbol s_n))^\top$ follows a multivariate normal distribution $\boldsymbol w\mid\sigma^2,\boldsymbol\theta\sim\mathcal N(\boldsymbol0,\boldsymbol K)$, where $\boldsymbol K$ is an $n \times n$ covariance matrix whose entries are $K(\boldsymbol s_i,\boldsymbol s_j)$. Let $\boldsymbol\eta:=(\eta(\boldsymbol s_1),\ldots,\eta(\boldsymbol s_n))^\top$ denote the vector of transformed site-specific conditional means, such that the data model is given by $g(\mathbb E[\boldsymbol z\mid\boldsymbol\beta, \boldsymbol w]):=\boldsymbol\eta=\boldsymbol X\boldsymbol\beta+ \boldsymbol w$ for some link function $g(\cdot)$. Here, each observation $z(\boldsymbol s_i)$ is assumed to be conditionally independent given $\eta(\boldsymbol s_i)$. Let $\boldsymbol X=[\boldsymbol x(\boldsymbol s_1),\ldots,\boldsymbol x(\boldsymbol s_n)]^\top$ denote the covariate matrix. Then under the Bayesian hierarchical framework, SGLMMs are structured as follows:
\begin{equation}\label{EQ:SGLMM}
\begin{aligned}
&\textbf{Data Model: } & z(\boldsymbol s_i)\mid\eta(\boldsymbol s_i)\overset{indpt}{\sim} F(\eta(\boldsymbol s_i)) \\
& &g(\mathbb E[\boldsymbol z\mid\boldsymbol\beta, \boldsymbol w]):=\boldsymbol\eta=\boldsymbol X\boldsymbol\beta+ \boldsymbol w\\
&\textbf{Process Model: } &\boldsymbol w\mid\sigma^2,\boldsymbol\theta\sim\mathcal N(\boldsymbol 0, \boldsymbol K)\\
&\textbf{Parameter Model: } & \sigma^2\sim p\left(\sigma^2\right), \boldsymbol\theta\sim p(\boldsymbol\theta),
\end{aligned}
\end{equation}
\noindent with prior distributions $p\left(\sigma^2\right)$ and $p(\boldsymbol\theta)$ specified by the practitioner.

Despite their flexibility, SGLMMs are subject to a myriad of limitations. In many cases, SGLMMs assume a second-order stationary and isotropic GP for the spatial random effects $\boldsymbol w$, where the covariance function depends solely on pairwise Euclidean distances. This assumption can be overly restrictive or unrealistic \citep{risser2016nonstationary}, especially for large and heterogeneous spatial domains \citep{katzfuss2013bayesian}. Furthermore, evaluating the density $\boldsymbol w\sim\mathcal N(\bm 0,\boldsymbol K)$ in SGLMMs requires $\mathcal O\left(n^3\right)$ operations and $\mathcal O\left(n^2\right)$ memory, which can be computationally prohibitive for large datasets. SGLMMs are often overparameterized, since all $n$ highly-correlated random effects must be inferred. This can result in slow-mixing Markov chains in the MCMC algorithm \citep{haran2011gaussian}. 

\subsection{Basis Function Representations}
\label{subsec:basisfunctions}

As discussed in the introduction, to alleviate the inferential and computational concerns of parameterizing all $n$ highly-correlated random effects, the spatial random effects can be modeled as a low-rank (dimension-reduced) process using a basis function expansion. Basis representation approaches have been widely used for modeling complex spatial processes due to their flexibility and computational efficiency \citep{cressie2011statistics, cressie2022basis,bradley2011selection}. These approaches represent the spatial process $w(\cdot)$ as a linear combination of $m$ basis functions $\boldsymbol\Phi(\boldsymbol s)=(\Phi_1(\boldsymbol s),\ldots,\Phi_m(\boldsymbol s))^\top$ with corresponding basis coefficients $\boldsymbol\delta=(\delta_1,\ldots,\delta_m)^\top$ such that,
\begin{equation}\label{EQ:basis}w(\boldsymbol s)\approx\sum_{i=1}^m \Phi_i(\boldsymbol s)\delta_i=\boldsymbol\Phi(\boldsymbol s)^\top\boldsymbol\delta,\quad \boldsymbol s\in\mathcal D,\end{equation}
\noindent where $\boldsymbol\delta\mid\boldsymbol\Sigma_{\boldsymbol\delta}\sim\mathcal N_m(\boldsymbol 0,\boldsymbol\Sigma_{\boldsymbol\delta})$. To reduce the dimension (or rank) and alleviate computational costs, we retain only the first $m\ll n$ basis functions. Let $\boldsymbol\Phi$ be an $n\times m$ matrix with columns indicating the basis functions and rows indicating the locations. Then by construction of (\ref{EQ:basis}), the approximated covariance matrix of $\boldsymbol w$ is $\boldsymbol\Phi\boldsymbol\Sigma_{\boldsymbol\delta}\boldsymbol\Phi^\top$. This does not solely depend on the distance between locations and is hence nonstationary. Furthermore, evaluating the density $\boldsymbol\delta\mid\boldsymbol\Sigma_{\boldsymbol\delta}\sim\mathcal N_m(\boldsymbol 0,\boldsymbol\Sigma_{\boldsymbol\delta})$ only involves matrix operations on matrices of size $m\times m$, which requires $\mathcal O\left(m^3\right)$ operations and $\mathcal O\left(m^2\right)$ storage.  

Different types of basis functions have been used, including radial basis functions, such as multi-resolution basis functions \citep{sengupta2016predictive,cressie2008fixed,katzfuss2011spatio,katzfuss2012bayesian}, Fourier basis functions \citep{xu2005kernel}, eigenfunctions \citep{holland1999spatial}, and the predictive-process approach \citep{banerjee2008gaussian}. Multi-resolution basis function approaches employ multiple layers of nested basis functions with varying resolutions to capture spatial structures from very fine to very large scale. For example, \cite{sengupta2016predictive} utilize a ``quad-tree'' structure comprised of low- and high-resolution bisquare basis functions. \cite{royle2005efficient} use Fourier basis functions to represent the spatial variability, a particularly useful method when working periodic or cyclical patterns in spatial data. When replicated spatial data are available, the eigenfunction approach employs eigenvectors of the dataset's empirical covariance matrix \citep{higdon2008computer,cressie2015statistics,mak2018efficient}, and the eigenvector basis functions capture the major patterns of spatial variation. When no replicates are available, a related approach extracts the $m$-leading eigenvectors of a proposed parametric covariance matrix (e.g., Mat\'ern class) \citep{guan2018computationally, christensen2006robust} that is dependent on pointwise distances. The covariance parameters can then be inferred within the modeling framework. The predictive-process approach considers both $\boldsymbol\Phi(\cdot)$ and $\boldsymbol\Sigma_w$ to be parameterized (e.g., Mat\'ern class) according to a lower-dimensional ``parent process.'' Given the ``parent process'' $w(\cdot)$, the predictive process is defined as, $w^*(\cdot)=\mathbb E(w(\cdot)\mid w(\boldsymbol u_1),\ldots,w(\boldsymbol u_m))$, where $\mathcal K=\{\boldsymbol u_1,\ldots,\boldsymbol u_m\}$ is a set of knots.

In this manuscript, we focus on radial basis functions (e.g. Gaussian, bisquare, thin-plate-splines), which are usually parameterized by knots and bandwidth parameters. Despite their flexibility and low costs, radial basis functions require the user to pre-specify the: (1) number of knots; (2) knot locations; and (3) bandwidth parameters. For example, \cite{sengupta2016predictive} fix the bandwidths and knots associated with each resolution of nested layers of bisquare basis functions. Similarly, \cite{nychka2015multiresolution} employ fixed compactly supported radial basis functions to capture multiple scales of spatial dependency. This pre-specification can potentially constrain the hierarchical spatial model to a fixed set of basis functions without any feedback or influence from the observed data. Given the challenge of appropriately specifying the number and placement of knots in spatial data, it is crucial to employ adaptive methods for knot selection. 

\begin{samepage}
\section{Adapt-BaSeS: Adaptive Basis Selection and Specification}
\label{sec:methodology}
\end{samepage}

\noindent Our proposed method, Adaptive Basis Selection and Specification (Adapt-BaSeS), is a flexible yet computationally efficient method for modeling nonstationary and non-Gaussian spatial data. The utility of Adapt-BaSeS comes from the adaptive tuning of radial basis functions, specifically the number and placement of knots as well as the bandwidths. Let $\mathcal K=(\boldsymbol u_1,\ldots,\boldsymbol u_m)$ be a vector of $m$ knots over the spatial domain $\mathcal D$ and let $\epsilon>0$ be the bandwidth parameter. Then the Gaussian radial basis function corresponding to knot $\boldsymbol u_i$ is defined as:
\begin{equation}\label{EQ:RadialBasis}
   \Phi_i(\boldsymbol s) = \exp(-(\epsilon||\boldsymbol s - \boldsymbol u_i||^2)),
\end{equation}
\noindent where $||\cdot||$ denotes the Euclidean distance between the points $\boldsymbol s$ \textit{and} $\boldsymbol u_i$. The choice of $\epsilon$ is crucial, as large values can lead to overfitting \citep{chaudhuri2017mean} and sharp localized peaks, while small values can oversmooth the latent spatial surface. Improper specification of the bandwidth parameters can lead to inaccurate predictions and improper approximations of the latent spatial surfaces \citep{sheather1991reliable,damodaran2018fast}. 

Adapt-BaSeS addresses these challenges by embedding an adaptive basis selection and specification mechanism within the SGLMM framework. First, we partition the spatial domain into disjoint subregions using an agglomerative clustering algorithm \citep{heaton2017nonstationary}. Next, we fit a hierarchical spatial model with partition-specific and adaptive radial basis functions to model the observed data. Our algorithmic approach employs a RJMCMC algorithm \citep{green1995reversible} to select key features of the radial basis functions (e.g., knot locations, total number of bases, and bandwidths) with clear feedback from the data. To the best of our knowledge, this study is the first to allow for both adaptive bandwidths and knots for basis-representation SGLMMs.

\subsection{Spatial Partitioning}
\label{subsec:partition}

\begin{figure}[h]
\begin{center}
\begin{tabular}{cc}
\includegraphics[width=70mm]{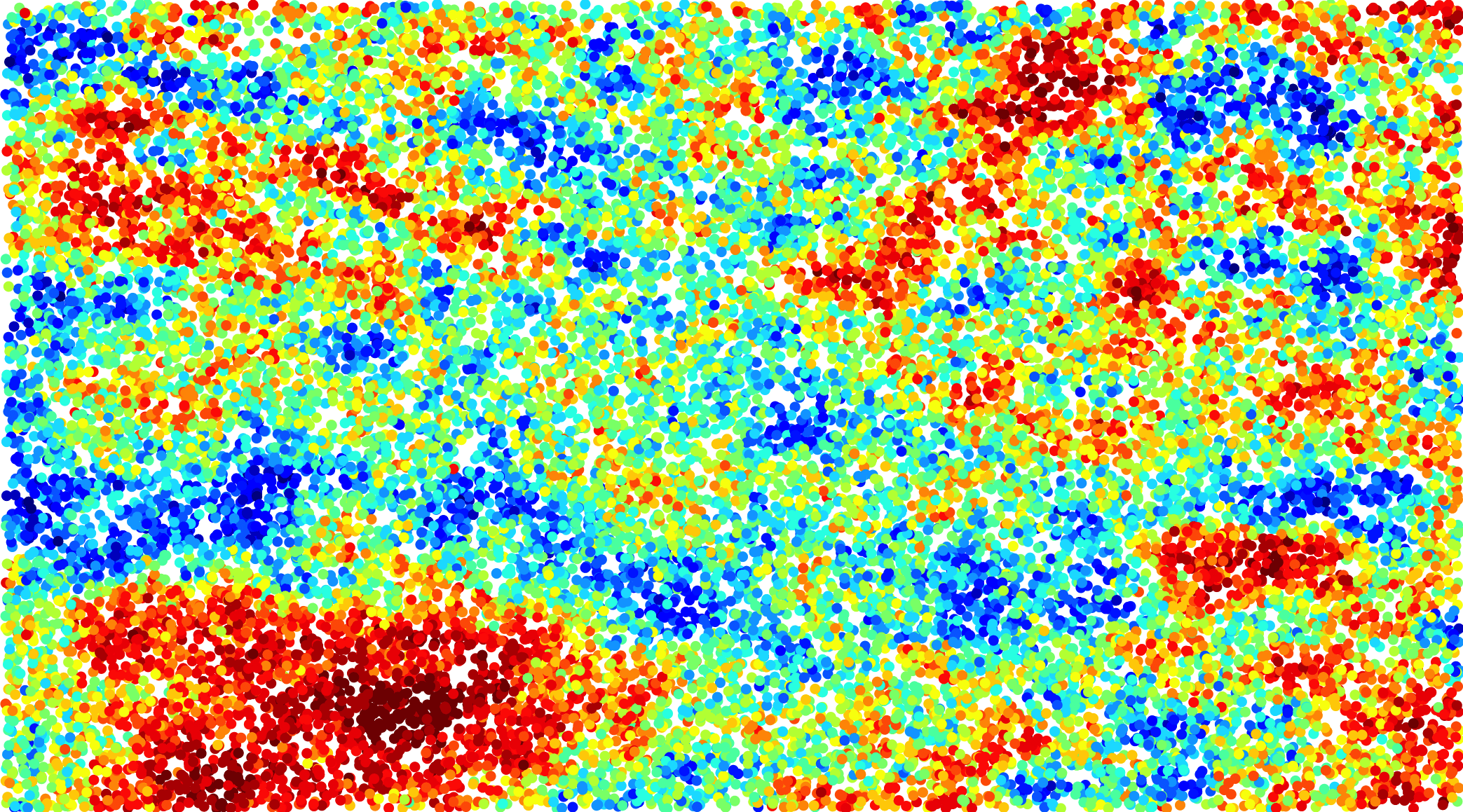} &   \includegraphics[width=70mm]{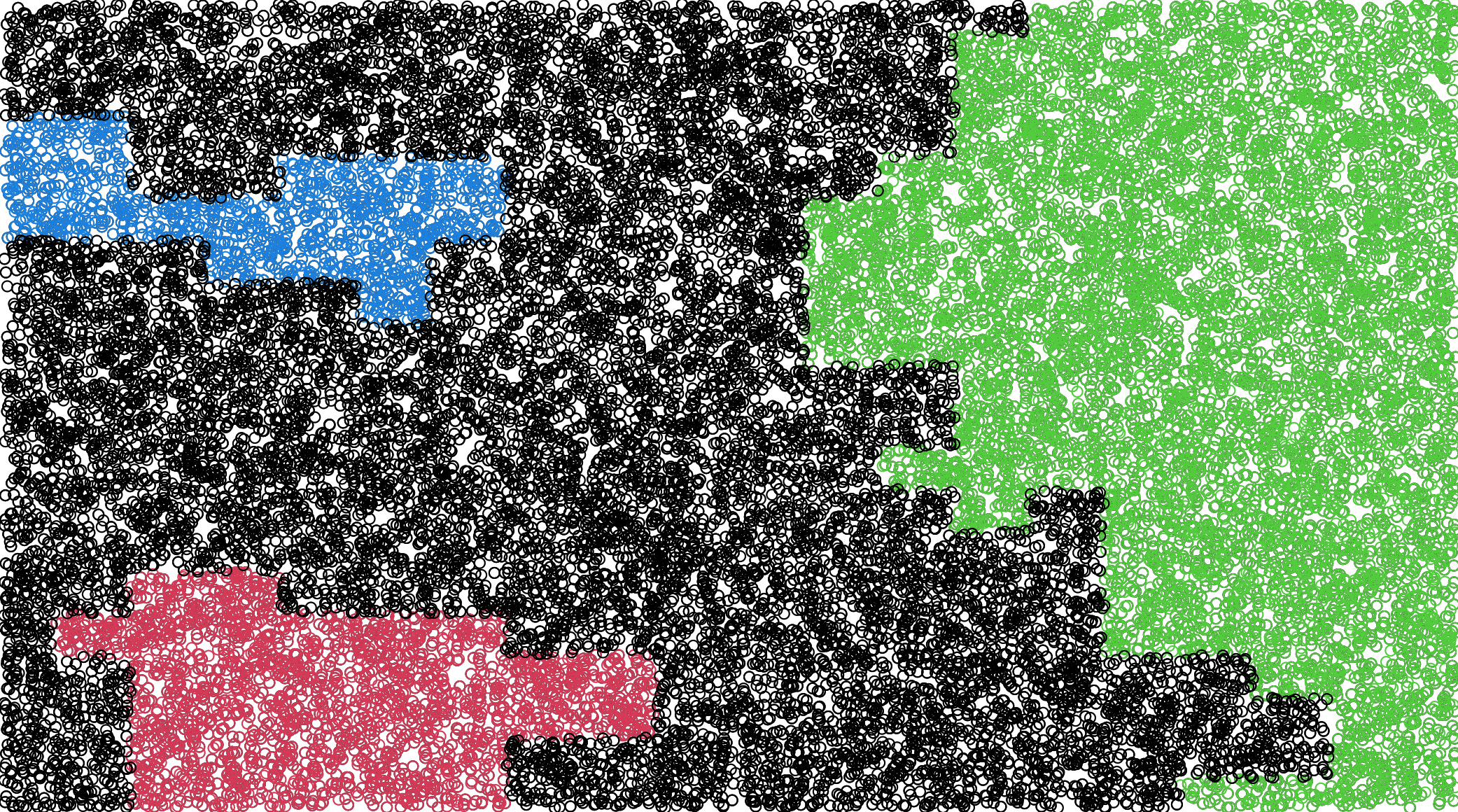} \\
(a) Spatial Surface & (b) Partition of Spatial Domain ($K=4)$ \\[12pt]
\includegraphics[width=70mm]{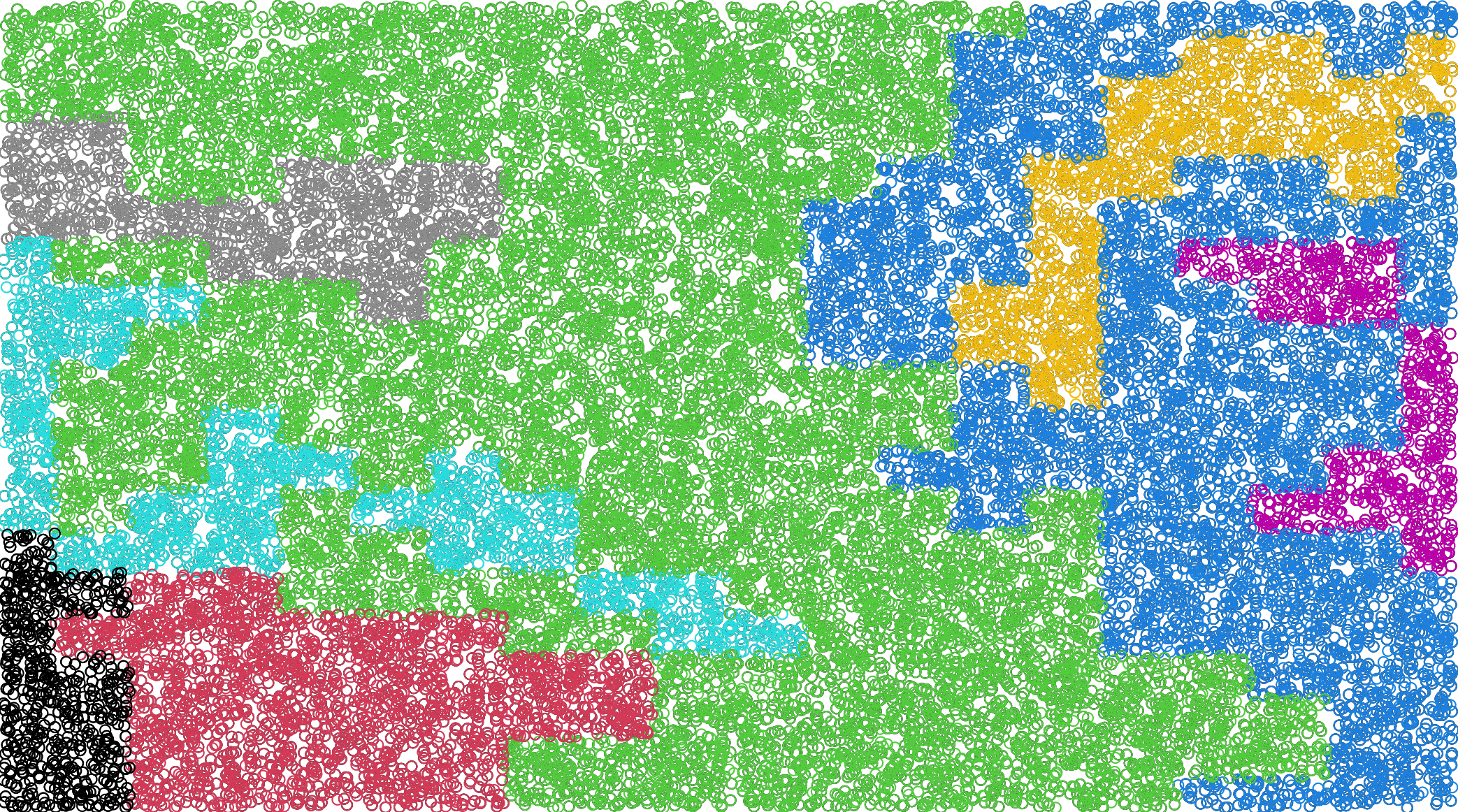} &   \includegraphics[width=70mm]{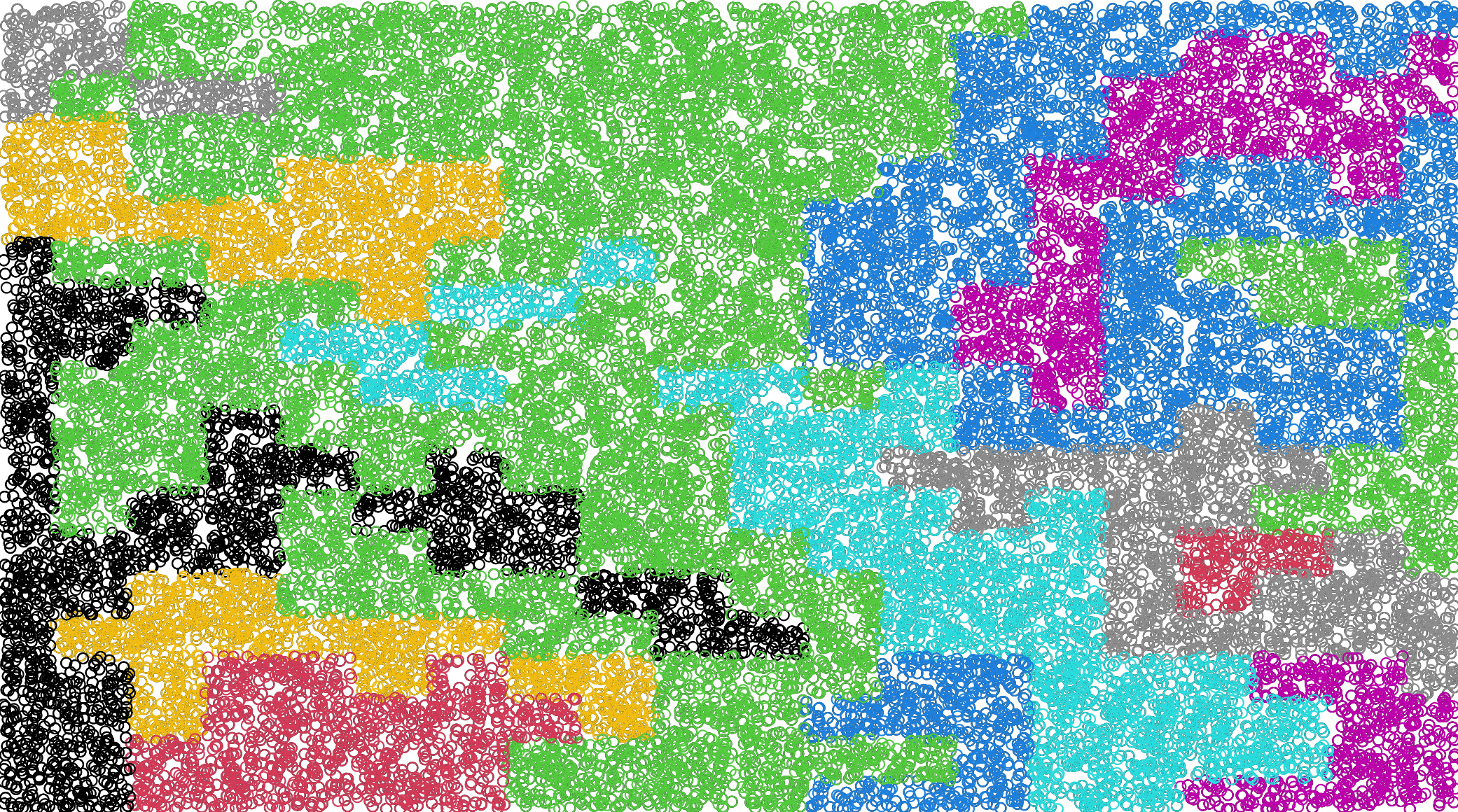} \\
(c) Partition of Spatial Domain ($K=8)$ & (d) Partition of Spatial Domain ($K=16)$ \\[2pt]
\end{tabular}
\caption{Illustration of spatial partitioning. (a) Spatial observations. (b) Observations partitioned into 4 subregions; different colors indicate disjoint partitions. (c) Case for 8 partitions. (d) Case for 16 partitions.}
\label{Fig1:Partitioning}
\end{center}
\end{figure}

Let $\boldsymbol z=(z(\boldsymbol s_1),\ldots, z(\boldsymbol s_n))^\top$ denote observations at locations $\boldsymbol s_i \in \mathcal D$. We use an agglomerative clustering approach \citep{heaton2017nonstationary} to partition
the spatial locations into $K$ disjoint subregions
$\{\mathcal D_k\}_{k=1}^K$ such that $\bigcup_{k=1}^K \mathcal D_k=\mathcal D$ and $\mathcal D_{i}\cap\mathcal D_{j}=\emptyset$ for all $i\neq j$. To accomplish this, the dissimilarity between $z(\boldsymbol s_i)$ and $z(\boldsymbol s_j)$ is:
$$d_{ij}=d(z(\boldsymbol s_i),z(\boldsymbol 
s_j))=\frac{\left|z(\boldsymbol s_i)-z(\boldsymbol s_j)\right|}{||\boldsymbol s_i-\boldsymbol s_j||}.$$
\noindent The dissimilarity metric $d_{ij}$ is based on spatial finite differences, which estimate the directional derivative \citep{banerjee2003directional}. This metric measures dissimilarity by considering how the spatial surface changes between two locations. When the surface changes rapidly in the direction from $\boldsymbol s_i$ to $\boldsymbol s_j$, the dissimilarity is large, indicating that $z(\boldsymbol s_i)$ and $z(\boldsymbol s_j)$ should be assigned to different clusters \citep{heaton2017nonstationary}. An agglomerative clustering approach is then used, where $K=n$ clusters are initialized such that each observation starts as its own cluster. Clusters are then linked together based on the smallest dissimilarity and spatial contiguity is enforced by only clustering Voronoi neighbors. This process is then repeated until the desired $K$ partitions is reached. We then let $\boldsymbol z_k=\{z(\boldsymbol s_i):\boldsymbol s_i\in\mathcal D_k\subset\mathcal D\}$ for $i=1,\ldots,n_k$ denote the $n_k$ observations belonging to the $k$-th partition, such that $n=\sum_{k=1}^K n_k$. For additional details on the clustering algorithm, please refer to the supplement. Alternatively, dissimilarity can be defined using the residuals from a regression with non-spatial errors \citep{heaton2017nonstationary}. In particular, the residuals are computed as $(z_i-\mathbb E[z_i\mid \boldsymbol x_i])$, where $\mathbb E[z_i\mid \boldsymbol x_i]$ comes from a generalized linear model. For count data (Poisson data model), $\mathbb E[z_i\mid \boldsymbol x_i]$ would be the expected intensity $\exp\left({\boldsymbol x^\top_i\hat{\boldsymbol{\beta}}}\right)$, and for binary data (Bernoulli data model), $\mathbb E[z_i\mid \boldsymbol x_i]$ would be the expected probability $1/\exp\left({-\boldsymbol x^\top_i\hat{\boldsymbol{\beta}}}\right)$. For non-Gaussian data, we find that this alternative approach outperforms clustering based on the observations themselves. Figure ~\ref{Fig1:Partitioning} illustrates the spatial partitioning for one simulated dataset, with different colors representing disjoint partitions.

\subsection{Bayesian Hierarchical Model}
\label{subsec:hierarchy}

For partition $k=1,\ldots, K$, the conditional mean $\mathbb E\left[\boldsymbol z_k\mid\boldsymbol\beta_k,\boldsymbol\delta_k,\epsilon_k,\mathcal K_k,\boldsymbol\gamma\right]$ is modeled as,
\begin{equation}\label{EQ:condInd}
\begin{gathered}
g\left(\mathbb E\left[\boldsymbol z_k\mid\boldsymbol\beta_k,\boldsymbol\delta_k,\epsilon_k,\mathcal K_k,\boldsymbol\gamma\right]\right):=\boldsymbol\eta_k=\boldsymbol X_k\boldsymbol\beta_k+\boldsymbol\Phi_k(\epsilon_k,\mathcal K_k)\boldsymbol\delta_k+\boldsymbol H_k\boldsymbol\gamma,
\end{gathered}
\end{equation}
\noindent where $\boldsymbol X_k$ is an $n_k\times p$ covariate matrix with corresponding regression coefficients $\boldsymbol\beta_k=(\beta_{k1},\ldots,\beta_{kp})^\top$, $\mathcal K_k=(\boldsymbol u_{k,1},\ldots,\boldsymbol u_{k,r_k})$ is a vector of partition-specific knots, $\epsilon_k$ is a partition-specific bandwidth parameter, $\boldsymbol\Phi_k(\epsilon_k,\mathcal K_k)$ is an $n_k\times r_k$ adaptive radial basis function matrix with basis coefficients $\boldsymbol\delta_k=(\delta_{k1},\ldots,\delta_{k r_k})^\top$, and $\boldsymbol H_k$ is a basis function matrix with global coefficients $\boldsymbol\gamma=(\gamma_1,\ldots,\gamma_g)^\top$. 

For the combined vector of observations $\boldsymbol z=(\boldsymbol z_1^\top,\ldots,\boldsymbol z_K^\top)^\top$, (\ref{EQ:condInd}) implies that,
$$g\left(\mathbb E\left[\boldsymbol z\mid\boldsymbol\beta,\boldsymbol\delta,\bm\epsilon,\mathcal K,\boldsymbol\gamma\right]\right):=\boldsymbol\eta=\boldsymbol X\boldsymbol\beta+\boldsymbol\Phi(\bm\epsilon,\mathcal K)\boldsymbol\delta+\boldsymbol H\boldsymbol\gamma,$$
\noindent where $\boldsymbol X$ is a block-diagonal matrix with matrices $\boldsymbol X_k$ on the main diagonal, $\boldsymbol\beta=(\boldsymbol\beta_1^\top,\ldots,\boldsymbol\beta_K^\top)^\top$, $\boldsymbol\epsilon=(\epsilon_1,\ldots,\epsilon_K)$, $\boldsymbol\Phi(\boldsymbol\epsilon,\mathcal K)$ is a block-diagonal matrix with matrices $\boldsymbol\Phi_k(\epsilon_k,\mathcal K_k)$ on the main diagonal, $\boldsymbol\delta=(\boldsymbol\delta_1^\top,\ldots,\boldsymbol\delta_K^\top)^\top$, and $\boldsymbol H$ is an $n\times g$ matrix stacking the individual $\boldsymbol H_k$ matrices.
The parameter $\epsilon_k$ determines the smoothness of the basis functions associated with the $k\text{-th}$ partition. By allowing $\epsilon_k$ to vary across partitions, our approach can capture the smooth and rough surfaces of the heterogeneous spatial domain. 

Since the basis functions (\ref{EQ:RadialBasis}) are infinitely differentiable, it is possible that the surfaces resulting from the basis expansions will be infinitely smooth. However, we find that letting $\epsilon_k$ be sufficiently large obviates this challenge in practical applications. Additionally, there may be concern that if the number of partitions $K$ is too small, the global and partitioned basis functions could be collinear due to their similar spatial scales. However, we have designed them to mitigate this issue. First, both types of basis functions have compact support. Second, the knots (basis centers) for the global and partitioned basis functions are unique and do not overlap. Third, the basis functions for the global and partitioned sets are from different classes. Specifically, we use bisquare basis functions for the global bases and Gaussian basis functions for the partitioned set. Fourth, since the partitioned basis functions are designed to capture finer-scale dependencies, we can limit the overlap by imposing constraints on the global basis functions' bandwidths. Similarly, we suggest practitioners use informative prior distributions for the partition-specific bandwidth parameters. Although we have made an effort to minimize collinearity issues, they may still arise in certain applications. In such cases, the model may become non-identifiable, compromising the estimation of the corresponding basis coefficients. However, it's worth noting that while this may affect the individual estimates of the basis coefficients, the overall predictions (kriging) of the model would remain sound, even if the individual estimates are not precise \citep{baker2022analyzing}.

In many applications, it may be desirable to have global regression coefficients such that $\boldsymbol\beta=\boldsymbol\beta_1=\cdots=\boldsymbol\beta_K$. In that setting, $\boldsymbol X$ would be an $n\times p$ matrix stacking the $\boldsymbol X_k$ matrices. Using Adapt-BaSeS, the hierarchical spatial model is:
\begin{equation}\label{OurMethod}
\begin{aligned}
&\textbf{Data Model: } &z(\boldsymbol s_i)\mid\eta(\boldsymbol s_i)\overset{indpt}{\sim} F(\eta(\boldsymbol s_i))\\
& &g(\mathbb E[\boldsymbol z\mid\boldsymbol\beta,\boldsymbol\delta,\boldsymbol\epsilon,\mathcal K,\boldsymbol\gamma]):=\boldsymbol\eta=\boldsymbol X\boldsymbol\beta+\boldsymbol\Phi(\boldsymbol\epsilon,\mathcal K)\boldsymbol\delta+\boldsymbol H\boldsymbol\gamma\\
&\textbf{Process Model: } &\boldsymbol\delta_k \mid\tau_k^2\sim\mathcal N(\boldsymbol 0, \tau_k^2\mathcal I)\\
 &&\boldsymbol\gamma \mid\rho^2\sim\mathcal N(\boldsymbol 0, \rho^2\mathcal I)\\
&\textbf{Parameter Model: } & \boldsymbol\beta_k\sim p(\boldsymbol\beta_k), \epsilon_k\sim p(\epsilon_k), \tau_k^2\sim p(\tau_k^2),\\&& \mathcal K_k\sim p(\mathcal K_k), \rho^2\sim p(\rho^2),
\end{aligned}
\end{equation}

\noindent  where $\mathcal I$ denotes the identity matrix. We complete the hierarchical spatial model by specifying the prior distributions for the model parameters $\boldsymbol\beta_k$, $\epsilon_k$, $\tau_k^2$, $\rho^2$, and $\mathcal K_k$. We can reduce computational costs by limiting the number of basis functions $r_k$ within each partition to be small and by specifying the covariance matrices $\boldsymbol\Sigma_{\boldsymbol\delta_k}$ and $\boldsymbol\Sigma_{\boldsymbol\gamma}$ to be diagonal \citep{lee2023scalable, higdon1998process,lindgren2011explicit,nychka2015multiresolution}. One can introduce a predefined covariance structure for the basis coefficients, such as employing an exponential covariance function \citep{heaton2019case}. However, this would result in additional computational overhead. Conditional on $\boldsymbol\gamma$, the parameters $\boldsymbol\beta_k$, $\mathcal K_k$, $\boldsymbol\delta_k$, $\epsilon_k$, and $\tau_k^2$ can be estimated independently for each partition. Hence, the MCMC updates of these parameters can be done in parallel to facilitate computational efficiency. 

We note that our approach assumes local nonstationarity because the partition-specific basis functions yield a nonstationary covariance structure. In addition, each partition has a different nonstationary covariance structure. This is in contrast to weighted-average methods approaches \citep{fuentes2001high,risser2015local,kim2005analyzing} which construct nonstationary spatial processes by smoothing locally stationary processes.

\subsection{The Reversible-Jump MCMC Algorithm}
\label{subsec:RJMCMC} 

We propose a RJMCMC algorithm (Algorithm~\ref{alg:RJMCMC}) to select the number and placement of knots within each partition. For a given partition $k$, we take the number of knots $r_k$ to be random, from some countable set $\mathcal S_k= \{0,\ldots, R_k\}$, where $R_k$ is the number of candidate knots in partition $k$. Let $\mathcal M_{r_k}$ denote the model with exactly $r_k$ knots and let $\mathcal K_{k}(r_k) = \{\boldsymbol u_{k,1},\ldots, \boldsymbol u_{k,r_k}\}$ denote the knots. We generate samples from the joint posterior of $(r_k, \mathcal K_{k}(r_k))$. To account for the varying dimensionality, we must develop appropriate reversible jump moves. For this problem, possible transitions are: (1) add a knot (birth step), (2) delete a knot (death step), and (3) move a knot. These independent move types are randomly chosen with probability $b_{r_k}$ for the move $r_k$ to $r_k +1$ (birth step), $d_{r_k}$ for the move $r_k$ to $r_k -1$ (death step), and $\eta_{r_k}$ for the move step. These probabilities must satisfy $b_{r_k} + d_{r_k} + \eta_{r_k} = 1$. For this choice, we define $b_0=d_{R_k}=1$ and  $b_{r_k}=d_{r_k}=\eta_{r_k}=1/3$ otherwise. 

\subsubsection{Prior Specifications}
\label{subsec:prior} 

We specify a truncated Poisson prior distribution for $r_k$, such that
$$p(r_k)\propto\frac{\lambda^{r_k} \exp(-\lambda)}{r_k!}\mathds{1}_{\{0,\ldots,R_k\}}(r_k).$$
\noindent The choice of $\lambda$ is a compromise between model flexibility and model parsimony. A small value of $\lambda$ reflects a strong incidence of smoothness whereas a large value may cause the model to fit the data too closely.  The choice of $\lambda$ will be discussed later.

For a given $r_k$, the knots are randomly selected with equal probability from a set of candidate knots $\mathcal{R}_k = \{\boldsymbol{u}_{k,1}, \ldots, \boldsymbol{u}_{k,R_k}\}$, where $\boldsymbol{u}_{k,1}, \ldots, \boldsymbol{u}_{k,R_k}$ are equally-spaced over the subregion $\mathcal{D}_k$. Given $r_k$, the prior distribution for $\mathcal K_{k}(r_k) = \{\boldsymbol u_{k,1},\ldots, \boldsymbol u_{k,r_k}\}$ is then given by
 $$p(\mathcal K_{k}(r_k)\mid r_k)={\binom{R_k}{r_k}}^{-1}=\frac{r_k!(R_k-r_k)!}{R_k!}.$$

A commonly used prior for regression coefficients of a generalized linear model is the multivariate normal distribution $\boldsymbol\delta_k\mid r_k\sim\mathcal N(\mathbf 0,\boldsymbol\Sigma_{\boldsymbol\delta_k})$ \citep{gamerman1997sampling}. Following \cite{biller2000adaptive}, we assume the basis coefficients $\boldsymbol\delta_k$ are uncorrelated, i.e., $\boldsymbol\Sigma_{\boldsymbol\delta_k}=\tau_k^2\mathcal I$. 

\subsubsection{Algorithm}
\label{subsec:alg} 

At each step of the RJMCMC algorithm, we propose one of three modifications to the current set of $r_k$ knots for each partition $k$:

 \begin{enumerate}
     \item \textbf{Add a knot (birth step):} Draw a new knot $\boldsymbol u_{k,r_k+1}$ uniformly with probability $1/(R_k- r_k)$ from the set of the $R_k - r_k$ vacant knots. Let $\mathcal K_k^*= \mathcal K_k\cup \{\boldsymbol u_{k,r_k+1}\}$ be the proposed set of knots, which now has size $r_k^* = r_k + 1$. 
     \item \textbf{Delete a knot (death step):} Select a knot $\boldsymbol u_{k,J}$ uniformly at random from one of the $r_k$ current knots, so it is drawn with probability $1/r_k$. Then set $\mathcal K_k^*= \mathcal K_k \backslash \{\boldsymbol u_{k,J}\}$ and $r_k^* = r_k - 1$. 
     \item \textbf{Move a knot (move step):} Select a knot $\boldsymbol u_{k,J}$ uniformly at random to be deleted, and then select a new location $\boldsymbol u_{k,r_k+1}$ from the vacant knots (i.e., where to move the old knot). This results in $\mathcal K_k^* = \{\boldsymbol u_{k,r_k+1}\}\cup\mathcal K_k\backslash\{\boldsymbol u_{k,J}\}$ and $r_k^* = r_k$.
 \end{enumerate}

\noindent Note that when we propose to add a knot $\boldsymbol u_{k,r_k+1}$, a corresponding basis coefficient $\delta_{k,r_k+1}^*$ will also need to be proposed. Similarly, if we propose to delete a knot $\boldsymbol u_{k,J}$, the current basis coefficient $\delta_{k,J}$ will be set to $0$. If we propose to move a knot, we will propose changing the current basis coefficient from $\delta_{k,J}$ to $\delta_{k,r_k+1}^*$. Complete details of the RJMCMC algorithm can be seen in Algorithm \ref{alg:RJMCMC}. Proposition \ref{proposition} asserts that a sufficient choice for the acceptance probability is given by (\ref{eq:acceptancebirth}); hence fulfilling the detailed balance condition. 

\begin{proposition}\label{proposition}
The detailed balance condition is satisfied by setting the acceptance probability to be $\min\{1,\alpha\}$, where
\begin{equation}\label{eq:acceptancebirth}
\alpha=\frac{L(\boldsymbol z_k\mid\mathcal K_k^*,\boldsymbol\delta_k^*)}{L(\boldsymbol z_k\mid\mathcal K_k,\boldsymbol\delta_k)}\frac{\pi(\mathcal K_k^*,\boldsymbol\delta_k^*)}{\pi(\mathcal K_k,\boldsymbol\delta_k)}\frac{Q(\mathcal K_k^*,\mathcal K_k)}{Q(\mathcal K_k,\mathcal K_k^*)},
\end{equation}
\noindent and the proposal ratio is given by, 
$$\frac{\mathcal Q(\mathcal K_k^*,\mathcal K_k)}{\mathcal Q(\mathcal K_k,\mathcal K_k^*)}=
\begin{cases}
\frac{R_k-r_k}{r_k+1}\times\frac{1}{\mathcal N(\delta_{k,r_k+1}^*;0,\sigma^2)},&r_k^*=r_k+1\\
\frac{r_k}{R_k-r_k+1}\times\mathcal N(\delta_{k,J};0,\sigma^2),&r_k^*=r_k-1\\
\frac{\mathcal N(\delta_{k,J};0,\sigma^2)}{\mathcal N(\delta_{k,r_k+1}^*;0,\sigma^2)},&r_k^*=r_k,
\end{cases}$$
\noindent and the prior ratio is given by, 
$$\frac{\pi(\mathcal K_k^*,\boldsymbol\delta_k^*)}{\pi(\mathcal K_k,\boldsymbol\delta_k)}=\begin{cases}
\frac{p(r_k+1)}{p(r_k)}\frac{r_k+1}{R_k-r_k}\frac{\mathcal N(\delta_{k,r_k+1}^*;0,\tau_k^2)}{\mathcal N(0;0,\tau_k^2)},&r_k^*=r_k+1\\ 
\frac{p(r_k-1)}{p(r_k)}\frac{R_k-r_k+1}{r_k}\frac{\mathcal N(0;0,\tau_k^2)}{\mathcal N(\delta_{k,J};0,\tau_k^2)},&r_k^*=r_k-1\\
\frac{\mathcal N(\delta_{k,r_k+1}^*;0,\tau_k^2)}{\mathcal N(\delta_{k,J};0,\tau_k^2)},&r_k^*=r_k,
\end{cases}$$

\noindent where $\sigma^2$ is the proposal variance for the basis coefficients. For the special cases where $r_k=0$ or $r_k=R_k$, the respective proposal ratios can be shown to be $R_k/(3\mathcal N(\delta_{k,r_k+1}^*;0,\sigma^2))$ and $(R_k\mathcal N(\delta_{k,J};0,\sigma^2))/3$.\\
Proof: See Supplement S5.
\end{proposition}

\begin{algorithm}
\caption{RJMCMC Algorithm}\label{alg:RJMCMC}
\begin{algorithmic}[1]
\For{$i \gets 1$ \textbf{to} $b$}
\State Metropolis-Hastings updates for $\boldsymbol\beta_k$, $\epsilon_k$, and $\boldsymbol\gamma$
\State Gibbs updates for $\tau_k^2$ and $\rho^2$
\For{Partition $k \gets 1$ \textbf{to} $K$}
\State Propose to: (1) add; (2) remove; or (3) move a knot with equal probability $1/3$
\State Denote proposed vector of knots and coefficients as $\mathcal K_k^*$  and $\boldsymbol\delta_k^*$, respectively

\State Generate $U\sim\text{Unif}(0,1)$ 
\State Accept $\mathcal K_k^*$ and $\boldsymbol\delta_k^*$ if $U<\min\{1,\alpha\}$ where $\alpha$ is defined in (\ref{eq:acceptancebirth}).
\EndFor
\EndFor
\end{algorithmic}
\end{algorithm}
\subsection{Prediction}
\label{subsec:prediction}

Let $n_o$ denote the number of observations used for model-fitting and let $n_p$ denote the number of observations used for validation. Upon fitting the hierarchical spatial model on the vector of observed locations $\mathcal S_o = (\boldsymbol s_1,\ldots, \boldsymbol s_{n_o})$, a natural extension is to infer the linear predictor $\eta(\cdot)$ at a vector of prediction locations $\mathcal S_p = (\boldsymbol s_1,\ldots, \boldsymbol s_{n_p})$. Letting $\boldsymbol s^*\in\mathcal S_p$ be an arbitrary unobserved location residing in partition $k$, we write: 
$$\eta(\boldsymbol s^*)=\boldsymbol x_k(\boldsymbol s^*)^{\top}\boldsymbol\beta_k+\boldsymbol\Phi_k(\boldsymbol s^*;\epsilon_k,\mathcal K_k)^{\top}\boldsymbol\delta_k+\boldsymbol H_k(\boldsymbol s^*)^\top\boldsymbol\gamma,$$
\noindent where $\boldsymbol x_k(\boldsymbol s^*)$ is the covariate vector evaluated at location $\boldsymbol s^*$, $\boldsymbol\Phi_k(\boldsymbol s^*;\epsilon_k,\mathcal K_k)$ is the adaptive basis function vector with corresponding basis coefficients $\boldsymbol\delta_k$, and $\boldsymbol H_k(\boldsymbol s^*)$ is the global basis function vector with corresponding global basis coefficients $\boldsymbol\gamma$. We approximate the posterior predictive distribution for $\eta(\boldsymbol s^*)$ using posterior samples $\{\boldsymbol\beta_k,\epsilon_k,\boldsymbol\delta_k,\mathcal K_k,\boldsymbol\gamma\}$. As with the parameter estimation, we note that the predictions are made within each cluster and hence can be done in parallel to promote computational efficiency. Furthermore, the posterior distribution allows for uncertainty quantification by evaluating the variance of the posterior samples.

\subsection{Implementation Details}

\noindent Our method requires tuning three parameters: (1) the number of partitions $K$; (2) the prior rate parameter $\lambda$ for the number of basis functions to be used within each partition; and (3) the prior distribution for the partition-specific bandwidths $\boldsymbol\epsilon=\{\epsilon_k\}_{k=1}^K$. Specifying fewer partitions (small $K$) may not be adequate for approximating nonstationary spatial processes because there may be several heterogeneous subregions within the spatial domain. Conversely, increasing the number of partitions (large $K$) may create multiple partitions containing very few observation locations (small $n_k$), potentially yielding a flat spatial surface. For practitioners, we suggest searching over a range of possible values. Specifically $K\in\{K_{\text{min}},\ldots,K_{\text{max}}\}$, where $K_{\text{min}}=\max\{4,\lceil\log_{10}(n)\rceil\}$, and $K_{\text{max}}$ is the largest value of $K$ such that each partition contains at least 50 observations. We set the minimum value of $K$ to be at least four because there are four global basis functions with the largest bandwidths. If cross validation is not feasible, one could use Akaike's information criterion (AIC), the Schwarz-Bayesian criterion (BIC), or adjusted $R^2$ from a model that includes the covariates, global basis functions, and cluster assignment as predictors. For instance, one can initially cluster observations using $K=\max\{4,\lceil\log_{10}(n)\rceil\}$ clusters and then incrementally increase $K$ either until the decrease in AIC is negligible (using the so-called ``elbow" method) or until $K_{\text{max}}$ is reached. An example of this procedure is provided in the supplement. In our simulation study, we compare the performance of our method with various choices of $K$. In fact, our method is fast enough such that practitioners can explore multiple $K$ settings and ultimately choose the most accurate $K$ based on out-of-sample predictions.3

For the partition-specific bandwidths, we specify a uniform prior $\epsilon_k\sim\text{Unif}(\alpha,\beta)$ to allow for control over the range of possible values $\epsilon_k$ can take on. For $\mathcal D=[0,5]^2$, a sensitivity analysis suggests that $\alpha=0.01$ and $\beta=3$ provides a suitable range of values for $\epsilon_k$ to accommodate both smooth and rough partitions. A truncated Poisson distribution with prior rate parameter $\lambda$ is used for the number of basis functions for each partition. A sensitivity analysis suggests that results are relatively insensitive to choices of $\lambda$ in the range 5-20 so we choose $\lambda=5$ to promote parsimony and computational efficiency. We set the priors for $\bm\beta$ and $\tau_k^2$ following \citep{hughes2013dimension}: $\boldsymbol\beta\sim\mathcal N(\boldsymbol 0,100\mathcal I)$ and $\tau_k^2\sim\mathsf{IG}(0.5,2000)$. The latter prior is desirable because it corresponds to the prior belief that the fixed effects are sufficient for explaining the data. For the global basis function matrix $\boldsymbol H$, we use three layers of nested bisquare basis functions \citep{sengupta2016predictive}. 

\section{Simulation Study}
\label{subsec:SimulationStudy}
In this section, we demonstrate the Adapt-BaSeS approach through an extensive simulation study featuring multiple spatial data classes and dependence structures. To benchmark performance, we compare our approach with two competing methods. 

\subsection{Simulation Study Design}
\label{subsec:SimulationStudyDesign}

Let $\boldsymbol s_i\in\mathcal D=[0,5]^2\subset\mathbb R^2$ for $i = 1,\ldots,n$ denote the spatial locations and let $\mathcal{\boldsymbol S} = (\boldsymbol s_1,\ldots,\boldsymbol s_n)$. On these locations, let $\boldsymbol z=(z(\boldsymbol s_1),\ldots, z(\boldsymbol s_n))^\top$ denote the vector of response variables (i.e., the data). For model fitting, we use $n_o=5{,}000$ observations and reserve $n_p=1{,}000$ observations for validation. We consider both binary and count data, with the associated spatial random effects generated from both nonstationary and stationary spatial processes. Observations are generated using the SGLMM framework described in (\ref{EQ:SGLMM}) with $x_1,x_2\sim\text{Unif}(-0.5,0.5)$ and $\boldsymbol\beta = (1,1)$. We study our method for $K = \{9, 16, 25, 36, 49\}$ partitions. All together, we study a total of $5\times2\times2 = 20$ implementations.

The nonstationary spatial random effects $\boldsymbol w = \{w(\boldsymbol s_i) :\boldsymbol s_i\in\mathcal D \}$ are generated by smoothing several locally stationary processes \citep{fuentes2001high}. Further details are provided in the supplementary material. The stationary spatial random effects are generated using an exponential covariance function with scaling parameter $\phi=1$ and partial sill parameter $\sigma^2=1$. The binary datasets use a Bernoulli data model and a logit link function and the count datasets are generated using a Poisson data model and a log link function. For each data generation mechanism (4 total), we generate 100 replicate data sets.

We fit the model using the hierarchical framework outlined in (\ref{OurMethod}). We generate $100{,}000$ samples from the posterior distribution $\pi(\boldsymbol\beta,\epsilon_k,\mathcal K_k, \boldsymbol\delta_k, \tau_k^2,\rho^2)$ for $k=1,\ldots, K$ using the RJMCMC algorithm described in Algorithm \ref{alg:RJMCMC}. To evaluate predictive performance, we compute the average root cross-validated mean squared prediction error ($\text{rCVMSPE}$), defined as $\text{rCVMSPE}=\left(\frac{1}{n_{p}}\sum_{i=1}^{n_{p}}\left(z_{i}-\hat{z}_{i}\right)^2\right)^{1/2}$, the area under the receiver operating curve (AUC) for the binary case, and the walltime (computation time) required to run $100{,}000$ iterations of the RJMCMC algorithm.

Fitting a ``gold standard'' SGLMM (\ref{EQ:SGLMM}) is prohibitive due to the overparameterized model and large matrix operations. Hence, we compare our approach with two competing methods: the NNGP approach and a fixed bisquare basis function approach. The computation times are based on a single 2.4 GHz Intel Xeon Gold 6240R processor provided by GMU's HOPPER high-performance computing infrastructure. 

\subsection{Simulation Study Results}
\label{subsec:SimulationStudyResults}

\begin{figure}[h]
\centering
\includegraphics[scale=.4]{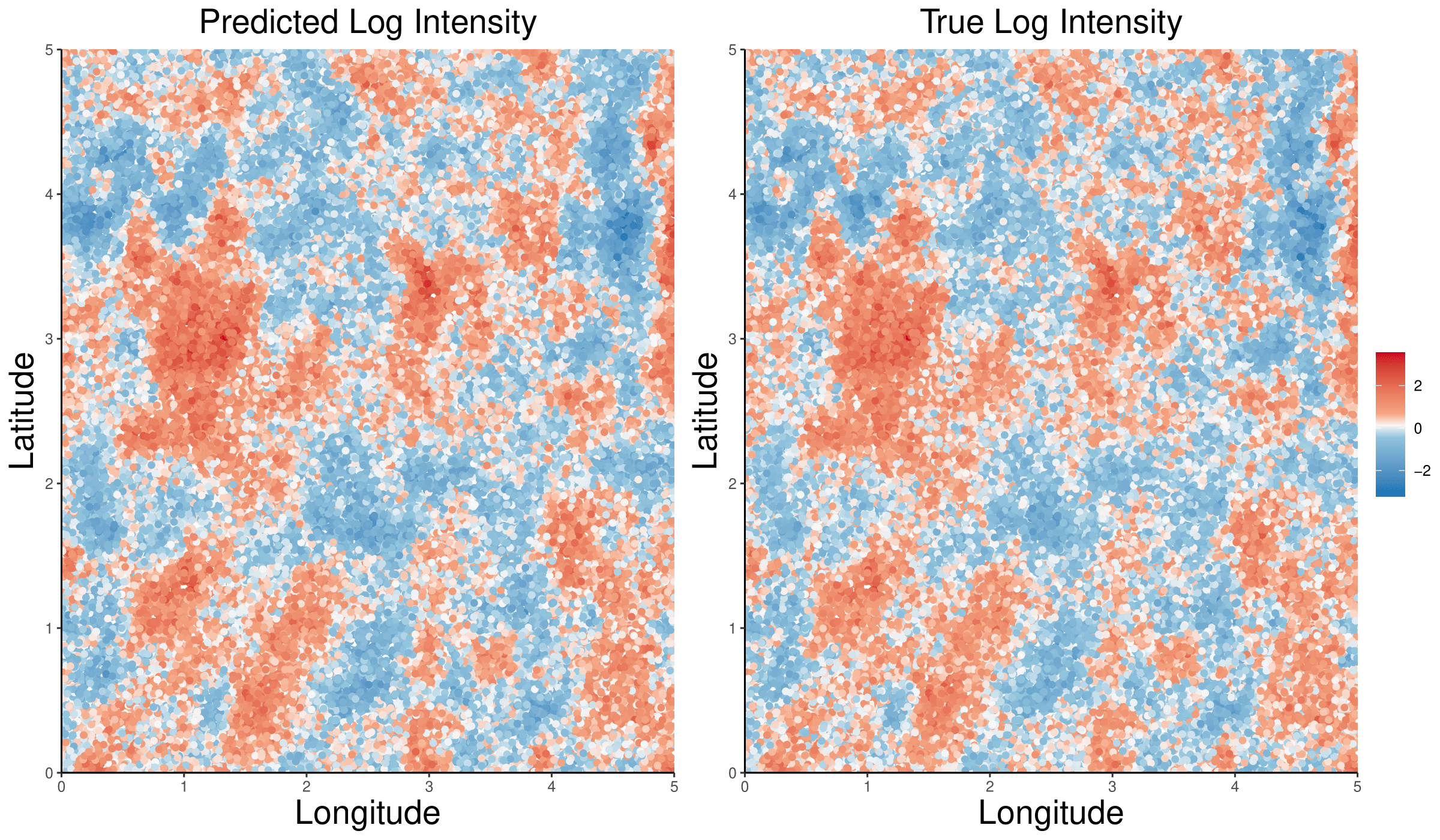}
\caption{Predicted (left) and true (right) log intensity surface for one nonstationary validation sample for the count data case.}
\label{Fig1:PredIntSurface}
\end{figure}

\begin{figure}[h]
\centering
 \includegraphics[scale=.4]{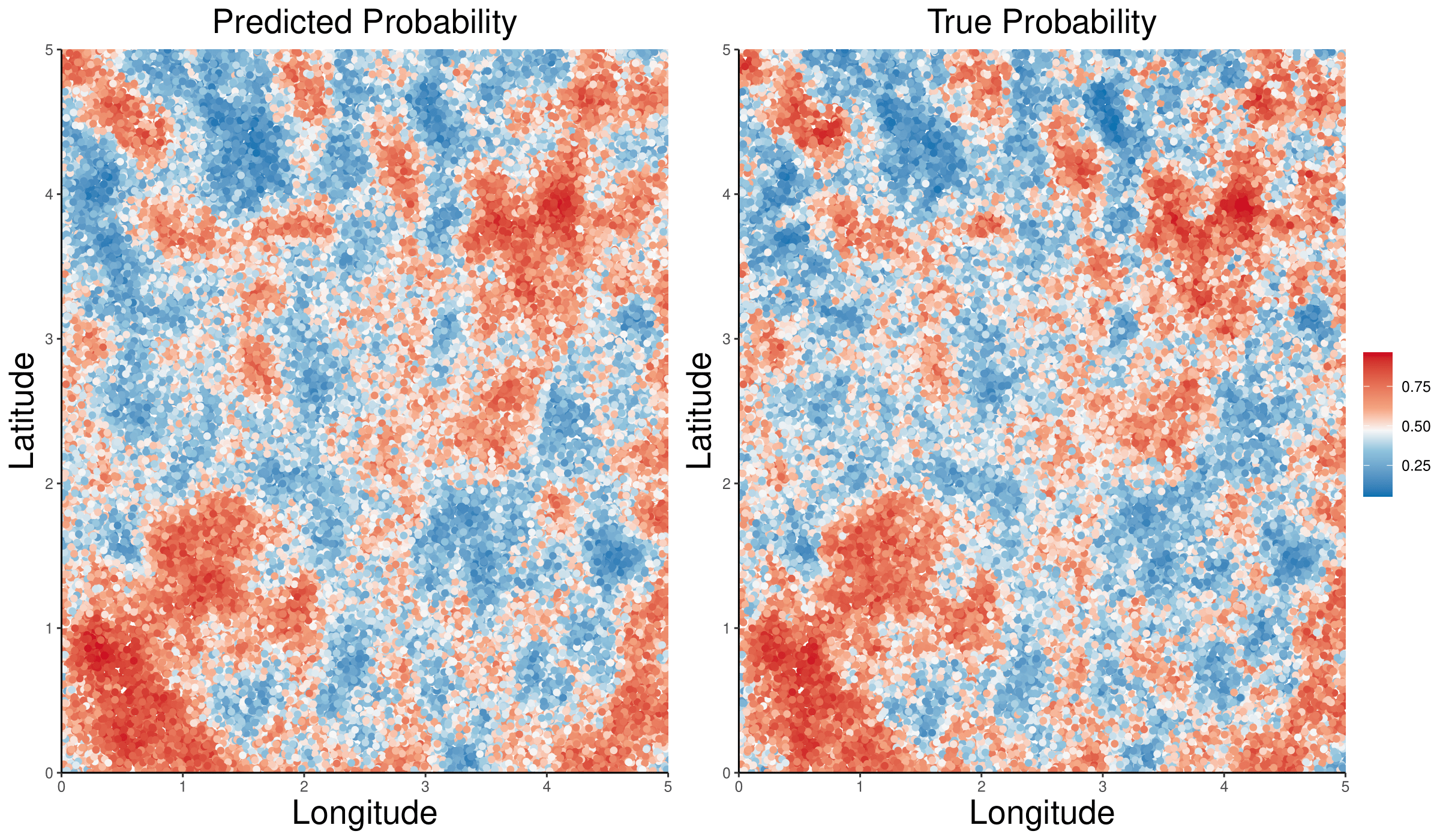}
\caption{Predicted (left) and true (right) probability surface for one nonstationary validation sample for the binary data case.}
\label{Fig3:PredProbSurface}
\end{figure}

\noindent Table~\ref{Tab1:SimulationStudy} and Table~\ref{Tab2:AUCSimStudy}, respectively, present the rCVMSPE and AUC for Adapt-BaSeS and the competing approaches. The results indicate that our method yields more accurate predictions than the competing methods across different values of $K$, data classes, and covariance structures. Paired $t$-tests were performed to compare the sets of rCVMSPE's between our approach and the competing methods. The corresponding $p$-values were found to be statistically significant, with $p<0.001$ for each pairwise comparison. Predictive performance generally improves as we increase the number of partitions $K$. However, the predictive standard deviations generally increase with larger $K$. For one simulated nonstationary dataset with 50{,000} observations, we present the posterior predictive log intensity surface (Figure \ref{Fig1:PredIntSurface}) and the posterior predictive probability surface (Figure \ref{Fig3:PredProbSurface}) obtained from the implementation yielding the lowest rCVMSPE ($K = 49$). Based on a visual inspection, our method successfully captures the nonstationary behavior of the true latent spatial process in both cases. Plots illustrating the prediction standard deviations, posterior differences, and coverage probabilities can be found in the supplementary material. The computation time required to fit the model to this large-scale dataset is 105 minutes on a laptop.

The model-fitting walltimes are reported in the supplementary material. The proposed approach exhibits higher computational costs than the fixed basis function approach. However, our method is more computationally efficient than the NNGP approach. The shorter walltimes for the fixed basis function approach are expected since the spatial basis functions are fixed prior to model-fitting. In contrast, our proposed method modifies the spatial basis functions at each iteration of the RJMCMC algorithm, leading to increased computational costs. Despite the longer walltimes, our approach offers additional flexibility in modeling the latent spatial process and yields more accurate predictions. Importantly, both approaches provide substantial improvements in computational efficiency over the ``gold standard'' SGLMM (\ref{EQ:SGLMM}), which would be computationally infeasible for a dataset with $n_o=5{,}000$ observations.

\begin{table}[h]
\begin{center} 
\begin{tabular}{cccccc} \toprule 
&\multicolumn{2}{c}{Nonstationary} && \multicolumn{2}{c}{Stationary}\\
\cmidrule{2-3}\cmidrule{5-6}
Method&Poisson & Binary && Poisson&Binary\\
\midrule
Bisquare & 1.817 (0.140) &0.478 (0.045)&& 1.754 (0.143) &0.460 (0.044)\\
\makecell{NNGP} & &0.474 (0.131)&&  &0.458 (0.123)\\
\arrayrulecolor{black!30}\midrule
$K=9$&1.719 (0.156) &0.470 (0.042)&& 1.709 (0.128)&0.453 (0.043) \\
$K=16$&1.700 (0.174) &0.468 (0.043)&& 1.650 (0.180)&0.451 (0.042) \\
$K=25$&1.688 (0.192) &0.466 (0.045)&& 1.648 (0.195)&0.450 (0.044) \\
$K=36$&1.682 (0.206)  &0.465 (0.048) && 1.639 (0.213)&0.449 (0.047) \\
$K=49$&1.680 (0.224)  &0.464 (0.052) &&1.633 (0.232) &0.448 (0.051)\\
\bottomrule \end{tabular}
\end{center}
\caption{Average rCVMSPE for the simulation study. Columns correspond to the data class and spatial dependence structure. Results are presented for various choices of $K$. Top rows correspond to the fixed basis functions approach and NNGP. Average standard deviation of the predictions are provided in parentheses.} 
\label{Tab1:SimulationStudy}
\end{table}

\begin{table}[h] 
\begin{center}
\begin{tabular}{ccc} \toprule 
Method & Nonstationary & Stationary \\ \midrule
Bisquare&0.663&0.711\\
\makecell{NNGP}&0.679&0.716\\
\arrayrulecolor{black!30}\midrule
$K=9$ & 0.693 & 0.731\\
$K=16$ & 0.700 & 0.736\\
$K=25$ & 0.706 &  0.740\\
$K=36$ & 0.709 &  0.742\\
$K=49$ & 0.711 &  0.743\\
\bottomrule \end{tabular} \end{center}
\caption{Out-of-sample area under the receiver operating curve (AUC)}
\label{Tab2:AUCSimStudy}
\end{table}

\section{Applications}
\label{subsec:Applications}

\noindent In this section, we apply the Adapt-BaSeS to two real-world spatial datasets: binary incidence of dwarf mistletoe in Minnesota \citep{hanks2011reconciling} and counts from the North American Breeding Bird Survey (BBS) \citep{BBS_2022}.

\subsection{Binary Data: Parasitic Infestation of Dwarf Mistletoe}
\label{subsec:Binary}

\noindent The dwarf mistletoe is a parasitic species that extracts key resources from its host, such as the black spruce species \citep{geils2002damage}. This infestation poses economic challenges, because black spruce is a valuable resource for producing high-quality paper. Thus, identifying regions with high probability of dwarf mistletoe being present would allow for targeted management and control efforts. We apply our method to analyze dwarf mistletoe incidence data in Minnesota, obtained from the Minnesota Department of Natural Resources operational inventory \citep{hanks2011reconciling}. The dataset contains binary incidence of dwarf mistletoe at $n=25{,}431$ locations, with dwarf mistletoe being present at $2{,}872$ of these locations. We fit the model on $n_o=12{,}931$ observations and set aside $n_p=12{,}500$ observations for validation. We consider several covariates as inputs to our model, including: (1) the average age of trees in years; (2) basal area per acre of trees in the stand; (3) average canopy height; and (4) volume of the stand measured in cords. We study the performance of our method for $K\in\{9,16,25,36,49\}$.

\begin{figure}[h]
\centering
\includegraphics[scale=.53]{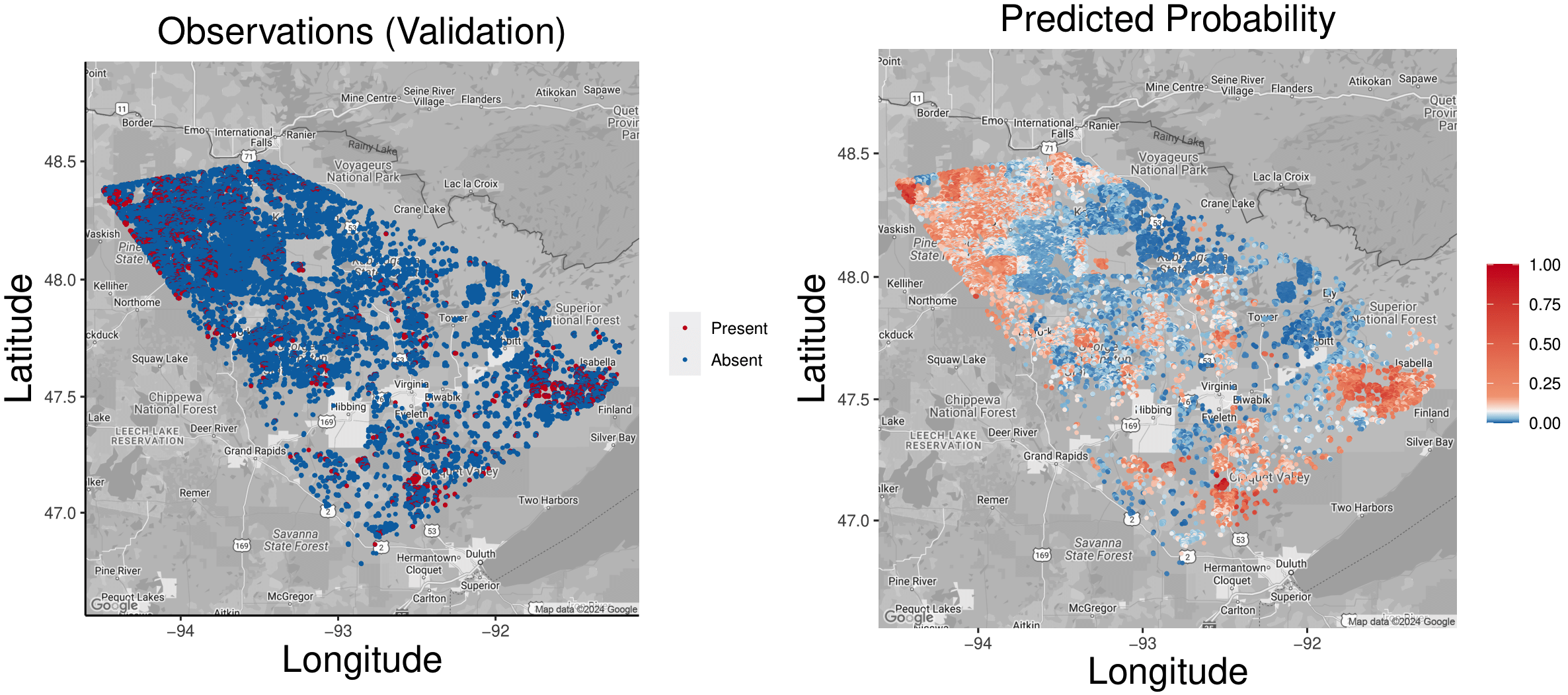}
\caption{Illustration of the dwarf mistletoe occurrence dataset for $K=49$. True observations (left) and posterior predictive probability surface (right) for the validation sample. Red points indicate the presence of dwarf mistletoe, while blue points indicate its absence (left plot). Similarly, red points indicate a high predicted probability of dwarf mistletoe, while blue points indicate a low predicted probability (right plot).}
\label{Fig:DwarfMistletoe}
\end{figure}

For each implementation, we compute the rCVMSPE and the AUC for the binary classification (Table~\ref{Tab3:AUCDwarf}). We observe that increasing the number of partitions improves the predictive performance of our proposed approach. Specifically, using $K = 49$ partitions yields the highest AUC value and the lowest rCVMSPE. In contrast, the fixed basis function approach provides less accurate predictions compared to our proposed method, across all five partition levels. Figure~\ref{Fig:DwarfMistletoe} displays the predictive probability surface and the true binary observations for the validation sample, for the case of $K=49$. The predictive probability map aligns with results from a previous study \citep{hanks2011reconciling} where dwarf mistletoe is abundant in the northwestern and middle eastern parts of northern Minnesota. Our results suggest that there are higher probabilities of dwarf mistletoe occurrence in these regions; hence, conservation hubs can be established there.

\begin{table}[h] 
\begin{center}
\begin{tabular}{ccc} \toprule 
Method & rCVMSPE & AUC \\ \midrule
Bisquare&0.308 (0.015) &0.734\\
\arrayrulecolor{black!30}\midrule
$K=9$ & 0.300 (0.017) & 0.764\\
$K=16$ & 0.296 (0.022) & 0.785\\
$K=25$ & 0.295 (0.023) & 0.790 \\
$K=36$ & 0.294 (0.024)& 0.793 \\
$K=49$ & 0.293 (0.025) & 0.795  \\
\bottomrule \end{tabular} \end{center}
\caption{rCVMSPE and AUC for the dwarf mistletoe dataset. Results are presented for varying number of partitions $K$. The top row corresponds to the fixed basis function approach. The average standard deviation of the predictions are provided in parentheses.}
\label{Tab3:AUCDwarf}
\end{table}

\subsection{Count Data: North American Breeding Bird Survey 2018}
\label{subsec:Count}

\noindent The Breeding Bird Survey (BBS) is an annual roadside survey that involves trained observers monitoring the abundance of bird populations in North America \citep{BBS_2022}. The bird surveys are conducted along more than 3,000 routes across the continental US, with 50 stops per route spaced roughly 0.5 miles apart. At each stop, observers conduct a three-minute point count where they record the total number of birds heard or seen \citep{pardieck2016}. We used the sum of counts from the 50 stops in one year's survey as an index of species abundance along the route for that specific year. Modeling species abundance is essential for informing resource management decisions, such as implementing sustainable harvesting practices or guiding habitat restoration efforts. For example, monitoring bird abundance has been fundamental to many successful programs aimed at studying and conserving bird populations \citep{brown2000united}. The particular BBS dataset includes Blue Jay (Cyanocitta cristata) bird counts at a total of $n=1{,}593$ locations, covering eastern and central regions of the United States. We use $n_o=1{,}000$ observations to fit the model and reserve $n_p=593$ for validation. The BBS data set does not provide any covariates, but environmental predictors can be obtained from various data sources. While this is a demonstration of our approach on a real-world dataset, future ecological studies can benefit from including covariates from multi-modal sources such as temperature, tree cover, and urban indices. We fit the model with only the spatial random effects (i.e., the conditional mean is modeled as $g(\mathbb E[z(\boldsymbol s_i)])=w(\cdot)$ and does not include spatial covariates). 

\begin{figure}[h]
\centering
\includegraphics[scale=.53]{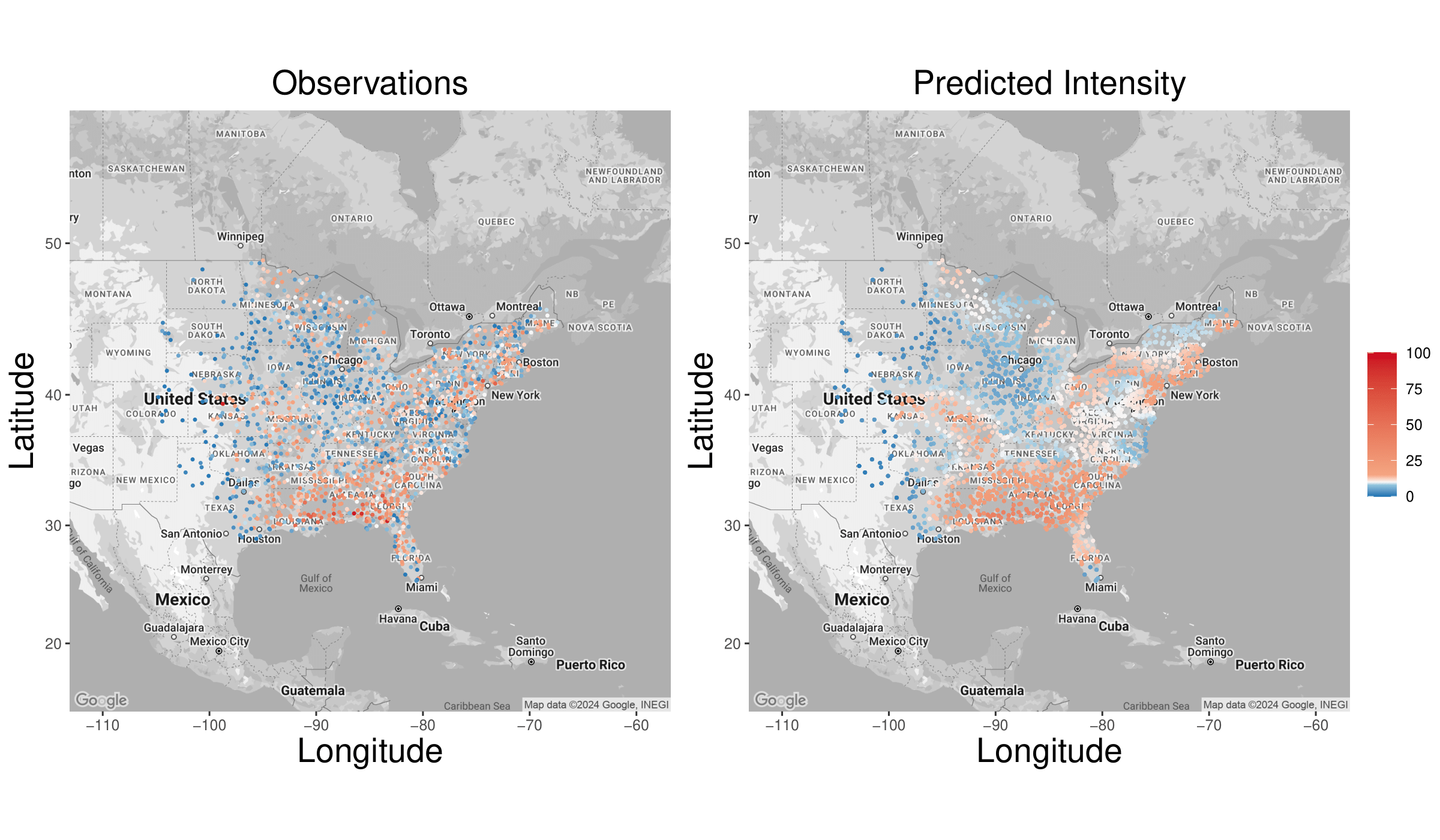}
\caption{Illustration of the Blue Jay count dataset for $K=8$. True observations (left) and posterior predictive intensity surface (right). Red points indicate a high abundance of Blue Jays, while blue points indicate a low abundance (left plot). Similarly, red points indicate a high predicted intensity of Blue Jays, while blue points indicate a low predicted intensity (right plot).}
\label{Fig4:BlueJay}
\end{figure}

Table \ref{Tab4:BlueJay} displays the rCVMSPE for each implementation. We find that our method consistently outperforms the fixed basis function approach. For the case of $K=8$ partitions, Figure \ref{Fig4:BlueJay} displays the true count observations and the predictive intensity surface, obtained from $k$-fold cross-validation. The predictive intensity map aligns with the observation of large counts in the southwestern and northeastern regions of the eastern United States. This pattern is consistent with the migratory behavior of Blue Jays, which migrate southwestward in the fall and northeastward in the spring in the eastern United States \citep{gill1941notes}. Our results can inform ecological and conservation policies, such as habitat protection and restoration, land-use planning, pesticide regulation, and curtailing the growth of invasive species. The higher intensity values in the northeast and southwestern regions of the eastern United States suggest that more resources should be allocated to the origin and destination of the migratory paths.

\begin{table}[h]
\begin{center} 
\begin{tabular}{cc} \toprule 
Method & rCVMSPE\\ \midrule
Bisquare&8.735 (0.706)\\
\arrayrulecolor{black!30}\midrule
$K=5$ & 8.599 (0.735)\\
$K=6$ & 8.583 (0.857)\\
$K=7$ & 8.537 (0.892) \\
$K=8$ & 8.515 (1.013)\\
$K=9$ & 8.572 (0.904) \\
\bottomrule \end{tabular} \end{center}
\caption{rCVMSPE for the Blue Jay spatial count dataset. Results are presented for varying $K$. Results from fixed basis functions are in the top row. Average standard deviation of predictions are in parentheses.}
\label{Tab4:BlueJay}
\end{table}

\section{Discussion}
\label{subsec:discussion}

We propose a data-informed, flexible, and computationally efficient method to model high-dimensional non-Gaussian spatial observations with nonstationary spatial dependence structures. Past studies have used spatial radial basis functions; thereby accounting for nonstationarity and reducing computational costs. However, these studies generally fix crucial components of the basis functions, such as the number of basis functions, placement of basis knots, and bandwidth (smoothing) parameters, perhaps arbitrarily, before fitting the model.

Our fast yet flexible method partitions the spatial domain into disjoint subregions using an agglomerative spatial clustering algorithm \citep{heaton2017nonstationary}. We then employ a RJMCMC algorithm to select critical features (knots and bandwidths) of the basis functions within each partition. Results from both our simulation study and real-world applications demonstrate that our approach performs well in both inference and predictions over competing methods, while also preserving computational efficiency. 

While our proposed adaptive framework primarily focuses on Gaussian radial basis functions, it can be extended to accommodate a wider range of radial basis functions. This includes thin-plate-spline basis functions, multiquadric radial basis functions, and bisquare basis functions, among others. Though our approach offers a significant speedup compared to the ``gold standard'' SGLMM (\ref{EQ:SGLMM}), the computational speedup can be further enhanced by embedding sparse basis functions such as the Wendland basis functions \citep{nychka2015multiresolution} or multi-resolution approximation (M-RA) basis functions \citep{katzfuss2017multi}, which can drastically reduce the number of floating point operations. In a similar strain, the embarrassingly parallel matrix operations can be distributed across available processors \citep{guan2018computationally} in high-performance computing systems.

The spatial statistics trinity refers to three key components that are fundamental to spatial analysis: population, sample, and inference \citep{wang2020spatial}. This study assumes that the data are collected from an unbiased sample and the population exhibits spatial autocorrelation. A biased sample can introduce biases in the model parameters and predictions \citep{ravishankar2023provable} and lead to less accurate estimates \citep{johnston2020estimating}. While our approach focuses on estimation, addressing sample conditions (such as biased samples) and population characteristics could be a natural extension for a future study. To address the error caused by a biased sample, existing approaches include creating synthetic data using imputation techniques in undersampled subregions or subsampling from oversampled subregions to make the sample more representative of the population \citep{li2017detecting}. Alternatively, employing weights can explicitly account for spatially biased presence-only datasets in species distribution modeling \citep{stolar2015accounting,johnston2020estimating}. Formal model-based approaches \citep{diggle2010geostatistical,conn2017confronting} jointly model the state variable of interest and the sampling site selection processes.

\section*{Supplementary Material}
The supplementary material includes: (1) details on the construction of the bisquare basis functions as well as a visualization of the multi-resolution ``quad-tree" structure; (2) details on the clustering algorithm \citep{heaton2017nonstationary}; (3) model-fitting walltimes for the simulation study; (4) a proof of the detailed balance proposition; (5) visualizations of the prediction standard deviation surfaces, posterior differences, and coverage probabilities; (6) a recommended procedure and demonstration for choosing the number of partitions $K$; (7) standard deviation surfaces for the real-world applications; and (8) inference results for the dwarf mistletoe data example.

\section*{Acknowledgements}\label{Sec:Acknowledgements}
The authors are grateful to Matthew Heaton, Murali Haran, Jaewoo Park, and Yawen Guan for providing helpful discussions as well as providing sample code. This project was supported by computing resources provided by the Office of Research Computing at George Mason University (\url{https://orc.gmu.edu}) and funded in part by grants from the National Science Foundation (Awards Number 1625039 and 2018631).


\end{document}


\doublespacing
\maketitle

\section{Reversible Jump MCMC}

Given the challenge of visually determining the appropriate number and placement of knots in spatial data, it is crucial to employ adaptive methods for knot selection. One such method is the reversible jump Markov chain Monte Carlo (RJMCMC) sampler \citep{green1995reversible}, which provides a flexible framework for Markov chain Monte Carlo simulation by allowing the dimension of the parameter space to vary at each iteration.

The RJMCMC sampler is particularly useful for model selection tasks, as it enables the Markov chain to explore parameter subspaces of different dimensions. This capability has led to successful application in various domains, including change-point analysis \citep{fan2000bayesian}, finite mixture models \citep{richardson1997bayesian}, time series models with an unknown number of components \citep{brooks2003efficient}, variable selection in regression models \citep{nott2004sampling}, and knot selection in curve fitting \citep{denison1998automatic}. 

Let $y$ be a vector of observations, and let $\mathcal M=\{\mathcal M_1,\mathcal M_2,\ldots\}$ denote a countable collection of candidate models, indexed by a model indicator $k\in\mathcal K$, for some countable set $\mathcal K$. Each model $\mathcal M_k$ has an $n_k$-dimensional vector of unknown parameters, $\theta_{k}\in\Theta_{k}\subset\mathbb R^{n_k}$, where $n_k$ can vary with $k$. The joint distribution of $(k,\theta_{k},y)$ is modeled as,
$$p(k,\theta_{k},y)=p(k)p(\theta_{k}\mid k) p(y\mid k,\theta_{k}),$$
\noindent where $p(k)$ is the model probability, $p(\theta_k\mid k)$ is the parameter prior given the model, and $p(y\mid k,\theta_k)$ is the likelihood. Inference about $k$ and $\theta_{k}$ is based on the joint posterior $p(k,\theta_{k}\mid y)\propto p(k,\theta_{k},y)$, which is known as the target distribution. For convenience, we abbreviate $(k,\theta_k)$ as $x$, and $p(k,\theta_k\mid y)=p(x\mid y)$ as $\pi(x)$. Given $k$, $x$ lies in $\mathcal C_k=\{k\}\times\Theta_k$, while generally, $x\in\mathcal C=\bigcup_{k\in\mathcal K}\mathcal C_k$. 

In order to traverse freely across the combined parameter space $\mathcal C$, we need a method that moves between parameter subspaces $\mathcal A,\mathcal B\subset\mathcal C$ of possibly different dimension. To that end, we consider different move types $m$, and for each of these move types we construct a transition kernel $P_m$, which satisfies the detailed balance condition,
\begin{equation}\label{EQ:detailedbalance}
 \int_{\mathcal A}\int_{\mathcal B}\pi(dx)P_m(x,dx')=\int_{\mathcal B}\int_{\mathcal A}\pi(dx')P_m(x',dx)   
\end{equation}
\noindent for all $\mathcal A,\mathcal B\subset\mathcal C$. This means that the equilibrium probability that the state of the chain is in a general set $\mathcal A$ and moves to a general set $\mathcal B$ is the same as with $\mathcal A$ and $\mathcal B$ reversed. To construct the MCMC sampler, consider $x=(k,\theta_k)$ to be the current state of the Markov chain. Following the Metropolis-Hastings algorithm, a move of type $m$ is proposed to a new state $x'=(k',\theta_{k'}')$ according to the proposal density $q_m(x,dx')$. As usual with the Metropolis-Hastings algorithm, the detailed balance condition (\ref{EQ:detailedbalance}) is enforced through the acceptance probability, where the move to the candidate state $x'$ is accepted with probability
$$\alpha_m(x,x')=\min\left\{1,\frac{\pi(dx')q_m(x',dx)}{\pi(dx)q_m(x,dx')}\right\},$$
\noindent otherwise, remain at the current state $x$. At each step of the RJMCMC algorithm, \cite{brooks2011handbook} divide the types of moves $m$ into two major categories:
\begin{itemize}
    \item {\it{Within-model moves:}} fix the model index $k$ and update the parameters $\theta_{k}$ following standard Gibbs or Metroplis-Hastings algorithms.
    \item {\it{Between-model moves:}} jointly update the state $x=(k,\theta_{k})$ by proposing a new state $x'=(k',\theta_{k'}')\sim q_m(x,dx')$ and ``matching dimensions" before accepting with probability $\alpha_m(x,x')$.
\end{itemize}
While the within-model moves are straightforward, the between-moves are more complicated as they involve a ``dimension matching" component \citep{green1995reversible}. Suppose that the current state $x=(k,\theta_k)$ has dimension $n_k$ under the current model $\mathcal M_k$, and the proposed state $x'=(k',\theta_{k'}')$ under model $\mathcal M_{k'}$ has dimension $n_{k'}$, where $n_k\neq n_{k'}$. In order to ``match dimensions" between the two model states, introduce an auxiliary variable for the transition $m$ from model $\mathcal M_k$ to model $\mathcal M_{k'}$ denoted $u\sim g_m$ and of dimension $r_m$ where the density $g_m$ is known. The new state $\theta_{k'}'$ is constructed as $(\theta_{k'}',u')=h_m(\theta_k,u)$ for some suitable deterministic function $h_m$. The reverse move from $x'$ to $x$ needs to be defined symmetrically by generating random numbers $u'$ of dimension $r_m'$ from the joint distribution $g_m'$ needed for the reverse move, to move from $\theta_{k'}'$ to $\theta_k$, using the inverse function $h_m'$ of $h_m$. In order for the transformation from $(\theta_k,u)$ to $(\theta_{k'}',u')$ to be a diffeomorphism, meaning the transformation and its inverse are differentiable, it must be the case that $n_k+r_m=n_{k'}+r_m'$.

When there are multiple possible moves $m$, we must generally also include the probability $j_m$ of choosing a specific move. The detailed balance requirement (\ref{EQ:detailedbalance}) can then be rewritten as
$$\label{EQ:detailedbalance2}
 \int_{\mathcal A}\int_{\mathcal B}\pi(x)j_m(x)g_m(u)\alpha_m(x, x')dx du=\int_{\mathcal B}\int_{\mathcal A}\pi(x')j_m(x')g_m'(u')\alpha_m(x', x)dx'du'$$
\noindent for all $\mathcal A,\mathcal B\subset\mathcal C$. Thus, a sufficient choice for the acceptance probability $\alpha_m$ corresponding to move type $m$ is given by 
$$\alpha_m(x,x')=\min\left\{1,\frac{\pi(x')j_m(x')g_m'(u')}{\pi(x)j_m(x)g_m(u)}\left|\frac{\partial(\theta_{k'}',u')}{\partial(\theta_k,u)}\right|\right\},$$
\noindent where the last factor corresponds to the Jacobian for the change of variable between $(\theta_k,u)$ and $(\theta_{k'}',u')$.

\section{Bisquare Basis Functions}
\setcounter{equation}{0}

\begin{figure}[ht]
\centering
\includegraphics[scale=.4]{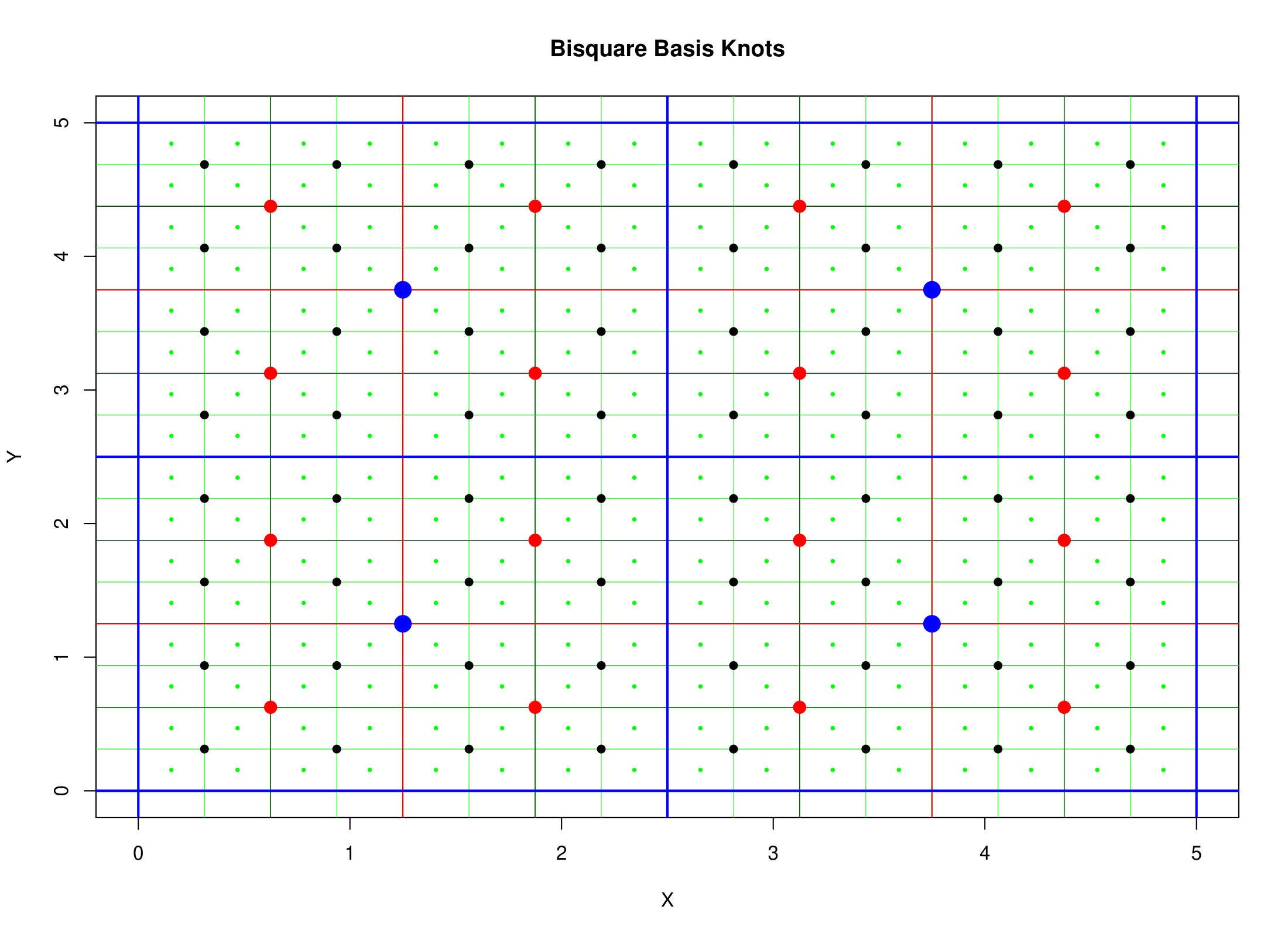}
\caption{Illustration of multi-resolution quad-tree structure}
\label{Fig1:bisquare}
\end{figure}

We employ the bisquare basis functions from \citep{sengupta2013hierarchical,cressie2008fixed} which take the form:
$$\Phi_m(\boldsymbol s)=\left\{1-\left(\frac{||\boldsymbol s-\boldsymbol u_m||}{\gamma}\right)^2\right\}^2\mathds{1}\left(||\boldsymbol s-\boldsymbol u_m||<\gamma\right),$$
\noindent where $\boldsymbol u_m$ is the center of basis function $m$ and $\mathds 1(\cdot)$ is an indicator function. The knots associated with each basis function are constructed according to a multi-resolution ``quad-tree'' structure such that the knots associated with different resolutions do not overlap. In particular, we use three resolutions, where there are four knot locations $\boldsymbol u_1,\ldots,\boldsymbol u_4$ for the first resolution, 16 knots $\boldsymbol u_5,\ldots,\boldsymbol u_{20}$ for the second resolution, and 64 knots $\boldsymbol u_{21},\ldots,\boldsymbol u_{84}$ for the third resolution. An illustration of the three resolutions of knot locations is provided in Figure \ref{Fig1:bisquare}. The bandwidth $\gamma$ for a specific resolution from \cite{cressie2008fixed} is given by $\gamma = 1.5\times\text{ minimum distance between knot locations}$.
We employ the multi-resolution bisquare basis functions into the Bayesian hierarchical framework for spatial generalized linear mixed models (SLGMMs) as follows:

$$\begin{aligned}
&\textbf{Data Model: } &z(\boldsymbol s_i)\mid\eta(\boldsymbol s_i)\sim F(\eta(\boldsymbol s_i)) \\
& &g(\mathbb E[\boldsymbol z\mid\boldsymbol\beta,\boldsymbol\delta]):=\boldsymbol\eta=\boldsymbol X\boldsymbol\beta+\boldsymbol\Phi\boldsymbol\delta\\
&\textbf{Process Model: } &\boldsymbol\delta \mid\tau^2\sim\mathcal N(\boldsymbol 0, \tau^2\mathcal I)\\
&\textbf{Parameter Model: } & \boldsymbol\beta\sim p(\boldsymbol\beta), \tau^2\sim p(\tau^2),
\end{aligned}$$

\noindent where $\mathcal I$ denotes the identity matrix and $\boldsymbol\Phi$ denotes the matrix of the multi-resolution bisquare basis functions. The hierarchical model is completed by assigning prior distributions for the model parameters $\boldsymbol\beta$ and $\tau^2$. In the simulation study, we use the following prior distributions for the model parameters: $\boldsymbol\beta\sim\mathcal N(\bm 0,100\mathcal I)$ and $\tau^2\sim\mathsf{IG}(0.5,2000)$.

\section{Spatial Clustering Algorithm}
\setcounter{equation}{0}

Here, we provide details on the clustering algorithm \citep{heaton2017nonstationary}. We obtain residuals $\bm{\epsilon}$ from a nonspatial generalized linear model (GLM) fit with a response vector $\boldsymbol{z}\in \mbR^{n}$ and a covariate matrix $\boldsymbol{X} \in \mbR^{n\times m}$. Let $\bm{\epsilon}_{k} \in \mbR^{n_{k}}$ denote the residuals belonging to the cluster (partition) $\mathcal{S}_k$. We can then define the dissimilarity between two clusters as 
\[
d(\mathcal{S}_{k_1},\mathcal{S}_{k_2})=\Big[\frac{N_{k_1}N_{k_2}}{N_{k_1}+N_{k_2}}(\bar{\epsilon}_{k_1}-\bar{\epsilon}_{k_2})^2\Big]\frac{1}{\bar{E}},
\]
where $\bar{\epsilon}_k$ is the average of the residuals in cluster $k$ and $\bar{E}$ is the average Euclidean distance between points in $\mathcal{S}_{k_1}$ and $\mathcal{S}_{k_2}$ amongst Voronoi neighbors. \cite{heaton2017nonstationary} define $\boldsymbol s_i$ and $\boldsymbol s_j$ to be Voronoi neighbors if they
share a border in a Voronoi tessellation of the observation locations $\boldsymbol s_1,\ldots\boldsymbol s_n$. The spatial clustering algorithm is summarized in Algorithm~\ref{clusteringalg}.

\begin{algorithm}
\caption{Spatial clustering algorithm \citep{heaton2017nonstationary} }\label{clusteringalg}
\begin{algorithmic}[ht]
\normalsize
\State Initialize $\mathcal{S}_k=\boldsymbol{s}_k$ for $k=1,\cdots,n$ such that each observation is its own cluster.

\State 1. Find clusters $\mathcal{S}_{k_1}$,$\mathcal{S}_{k_2}$ having the minimum
$d(\mathcal{S}_{k_1},\mathcal{S}_{k_2})$ where $\boldsymbol{s}_i \sim \boldsymbol{s}_j$ (Voronori neighbors) for $\boldsymbol{s}_i \in \mathcal{S}_{k_1}$ and $\boldsymbol{s}_j \in \mathcal{S}_{k_2}$

\State 2. Combine two clusters $$\mathcal{S}_{\min\lbrace k_1,k_2 \rbrace} = \mathcal{S}_{k_1} \cup \mathcal{S}_{k_2}$$ and set $$\mathcal{S}_{\max\lbrace k_1,k_2 \rbrace} =\emptyset$$

\State Repeat 1-2 until we have $K$ clusters where $K<n$.
\end{algorithmic}
\end{algorithm}

We note that Algorithm~\ref{clusteringalg} becomes computationally expensive when the number of observations is large. Following suggestions in \cite{heaton2017nonstationary}, we perform clustering after aggregating observations to a lattice $\lbrace \boldsymbol{s}^{\ast}_{l} \rbrace_{l=1}^{L}$ ($L<<n$). Here, $\mathcal{N}_{l}=\lbrace \boldsymbol{s}_i: \|\boldsymbol{s}_i-\boldsymbol{s}^{\ast}_{l}\| < \|\boldsymbol{s}_i-\boldsymbol{s}^{\ast}_{m}\|\text{ for all } l\neq m\rbrace$ is the subset of observations whose closest lattice point is $\boldsymbol{s}^{\ast}_{l}$, and $\bar{\epsilon}(\mathbf{s}^{\ast}_{l})= |\mathcal{N}_l|^{-1}\sum_{s_i \in \mathcal{N}_{l}} \epsilon(\boldsymbol{s}_{i})$ is the average of the observed residuals in $\mathcal N_l$. We then apply Algorithm ~\ref{clusteringalg} to $\lbrace \bar{\epsilon}(\boldsymbol{s}^{\ast}_{l}) \rbrace_{l=1}^{L}$ rather than to $\lbrace \epsilon(\boldsymbol{s}_{i}) \rbrace_{i=1}^{n}$. By specifying the number of lattice points $L$ to be much smaller than the number of observations $n$, the spatial clustering algorithm becomes computationally feasible. For example, in our simulation studies, we specify $L=400$ for $n=5{,}000$.

\section{Computation: Model-Fitting Walltimes}
\setcounter{equation}{0}
In Table ~\ref{Tab1:SimulationStudy}, we report the model-fitting walltimes (computation times). These include the time required for model initialization and running $100{,}000$ iterations of the RJMCMC algorithm. The proposed approach exhibits higher computational costs compared to the fixed bisquare basis function approach. However, our method is more computationally efficient than the NNGP approach. 

\begin{table}[h]
\begin{center} 
\begin{tabular}{cccccc} \toprule 
&\multicolumn{2}{c}{Nonstationary} && \multicolumn{2}{c}{Rough Stationary}\\
\cmidrule{2-3}\cmidrule{5-6}
Method&Poisson & Binary && Poisson&Binary\\
\midrule
\makecell{Fixed\\ (Bisquare)} & 192 &165&& 184&166\\
\makecell{NNGP} &  &2336&& &2291\\
\arrayrulecolor{black!30}\midrule
$K=9$&754 &581&& 726&602 \\
$K=16$&739 &633&& 748&643 \\
$K=25$&775 & 715&& 791&770 \\
$K=36$&833  & 830 && 860&842 \\
$K=49$& 931 &926 && 958& 932 \\
\bottomrule \end{tabular}
\end{center}
\caption{Average walltime (seconds) for 100,000 iterations}
\label{Tab1:SimulationStudy}
\end{table}

\section{Proof of Proposition}
\setcounter{equation}{0}

\noindent In this section, we provide a proof establishing that the acceptance probability satisfies the detailed balance condition for the birth step. Given the $r_k$ knots $\mathcal K_k=(\boldsymbol u_{k,1},\ldots,\boldsymbol u_{k,r_k})$, in the birth move we draw a new knot $\boldsymbol u^*$ uniformly with probability $1/(R_k- r_k)$ from the set of the $R_k - r_k$ vacant knot locations. Let $\mathcal K_k^*= \mathcal K_k\cup \{\boldsymbol u^*\}$ be the proposed set of knots, which now has size $r_k^* = r_k + 1$. The resulting model is now defined by the new model indicator $r_k+1$, the new knots $\mathcal K_k^*=(\boldsymbol u_{k,1}^*,\ldots,\boldsymbol u_{k,r_k+1}^*)$ (with $\boldsymbol u_{k,i}^*=\boldsymbol u_{k,i}$ for $i\leq r_k$ and $\boldsymbol u_{k,r_k+1}^*=\boldsymbol u^*$), and the new basis coefficients $\bm\delta_k^*=(\delta_{k,1}^*,\ldots,\delta_{k,r_k+1}^*)$, which must be adjusted appropriately. Formally, the birth move can be defined as a transition from state $\theta=(r_k,\theta_{r_k})$ to state $\theta^*=(r_k+1,\theta_{r_k+1}^*)$. With 
$$\theta_{r_k}=(\mathcal K_k,\boldsymbol \delta_k,\beta_k,\gamma,\tau_k^2,\rho^2,\epsilon_k)$$
\noindent and 
$$\theta_{r_k+1}^*=(\mathcal K_k^*,\boldsymbol\delta_k^*,\beta_k,\gamma,\tau_k^2,\rho^2,\epsilon_k),$$
\noindent there is a change in dimension from $\dim(\theta_{r_k})=2r_k+P+G+3$ to $\dim(\theta_{r_k+1}^*)=2r_k+P+G+5$. For the birth move, we have to compute $\theta_{r_k+1}^*$ as a function of $\theta_{r_k}$ and two random numbers $\boldsymbol u^*$ and $u$, with $u_B=(\boldsymbol u^*,u)$. The proposed knot $\boldsymbol u^*$ is drawn uniformly with probability $p(\boldsymbol u^*)=1/(R_k-r_k)$ from the set of the $R_k-r_k$ vacant candidate knots. The new coefficient $\delta_{k,r_k+1}^*$ corresponding to the new knot $\boldsymbol u_{k,r_k+1}^*=\boldsymbol u^*$ is set equal to $u$ where $u\sim g(u)$, i.e., we simulate a value of $u$ from some proposal distribution $g$. We specify $u\sim\mathcal N(0,\sigma^2)$ where $\sigma^2$ can be tuned by the user to achieve a reasonable acceptance rate. Hence, we set
\begin{align*}
  \delta_{k,r_k+1}^* &= u \\
  \delta_{k,r_k}^* &= \delta_{k,r_k} \\
        &\mathrel{\makebox[\widthof{=}]{\vdots}} \\
  \delta_{k,1}^* &= \delta_{k,1}
\end{align*}
\noindent This ensures that in the reverse death move, given the knot $\boldsymbol u_{k,r_k+1}^*$ to be deleted, the computation of $\bm\delta_k$ from $\bm\delta_k^*$ is deterministic and the required dimension matching holds:
\begin{align*}
  u &= \delta_{k,r_k+1}^* \\
  \delta_{k,r_k} &= \delta_{k,r_k}^* \\
        &\mathrel{\makebox[\widthof{=}]{\vdots}} \\
  \delta_{k,1} &= \delta_{k,1}^*
\end{align*}
\noindent The acceptance probability for the birth move can be expressed as
$$\alpha(\theta_{r_k},\theta_{r_k+1}^*)=\min\{1,\mathcal L\cdot\mathcal A\cdot\mathcal P\cdot\mathcal J\},$$
\noindent where $\mathcal L$ is the likelihood ratio, $\mathcal A$ is the prior ratio, $\mathcal P$ is the proposal ratio, and $\mathcal J$ is the Jacobian. The ratio of priors results in 
\begin{align*}
\mathcal A
&=\frac{\text{prior for }r_k+1\text{ knots}}{\text{prior for }r_k\text{ knots}}\times\frac{\text{prior for location of }r_k+1\text{ knots}}{\text{prior for location of }r_k\text{ knots}}\\
&\times\frac{\text{prior for proposed basis coefficient}}{\text{prior for current basis coefficient}}\\
&=\frac{p(r_k+1)}{p(r_k)}\frac{r_k+1}{R_k-r_k}\times\frac{\mathcal N(\delta_{k,r_k+1}^*;0,\tau_k^2)}{\mathcal N(0;0,\tau_k^2)},
\end{align*}
\noindent where the factor $p(r_k+1)/p(r_k)$ depends on the alternative priors of $r_k$: $p(r_k)\propto\frac{\lambda^{r_k} \exp(-\lambda)}{r_k!}\mathds{1}_{\{0,\ldots,R_k\}}(r_k)$ with prior rate parameter $\lambda$. The corresponding proposal ratio $\mathcal P$ is given by
\begin{align*}
\mathcal P
&=\frac{d_{r_k+1}(1/(r_k+1))}{b_{r_k}(1/(R_k-r_k))} \times\frac{1}{\mathcal N(\delta_{k,r_k+1}^*;0,\sigma^2)}\\
&=\frac{d_{r_k+1}(R_k-r_k)}{b_{r_k}(r_k+1)}\times\frac{1}{\mathcal N(\delta_{k,r_k+1}^*;0,\sigma^2)}.\\
\end{align*}
\noindent Considering $\theta_{r_k+1}^*$ as a function of $\theta_{r_k}$ and $u_B$, the Jacobian is
\begin{align*}
\mathcal J
&=\left|\frac{\partial\theta_{r_k+1}^*}{\partial(\theta_{r_k},u_B)}\right|\\
&= \begin{vmatrix}
      \frac{\partial\delta_{k,1}^*}{\partial\delta_{k,1}} & \frac{\partial\delta_{k,2}^*}{\partial\delta_{k,1}} & \frac{\partial\delta_{k,3}^*}{\partial\delta_{k,1}} & \cdots & \frac{\partial\delta_{k,r_k+1}^*}{\partial\delta_{k,1}} \\
      \frac{\partial\delta_{k,1}^*}{\partial\delta_{k,2}} & \frac{\partial\delta_{k,2}^*}{\partial\delta_{k,2}} & \frac{\partial\delta_{k,3}^*}{\partial\delta_{k,2}} &\cdots & \frac{\partial\delta_{k,r_k+1}^*}{\partial\delta_{k,2}} \\
     \frac{\partial\delta_{k,1}^*}{\partial\delta_{k,3}} & \frac{\partial\delta_{k,2}^*}{\partial\delta_{k,3}} & \frac{\partial\delta_{k,3}^*}{\partial\delta_{k,3}} &\cdots & \frac{\partial\delta_{k,r_k+1}^*}{\partial\delta_{k,3}} \\
      \vdots & \vdots & \vdots  & \ddots & \vdots \\
      \frac{\partial\delta_{k,1}^*}{\partial u} & \frac{\partial\delta_{k,2}^*}{\partial u} & \frac{\partial\delta_{k,3}^*}{\partial u} & \ldots & \frac{\partial\delta_{k,r_k+1}^*}{\partial u} \\
    \end{vmatrix}\\
&= \begin{vmatrix}
      1 & 0 & 0 & \cdots & 0 \\
      0 & 1 & 0 & \cdots & 0 \\
     \vdots & \vdots & \vdots  & \ddots & \vdots \\
      0 & 0 & 0 & \cdots & 1 \\
    \end{vmatrix} \\
&=1.
\end{align*}

\noindent The proof for both the move and death steps follows similarly. 

\section{Standard Deviation Surfaces}
\setcounter{equation}{0}

\begin{figure}[H]
\centering
\includegraphics[scale=.4]{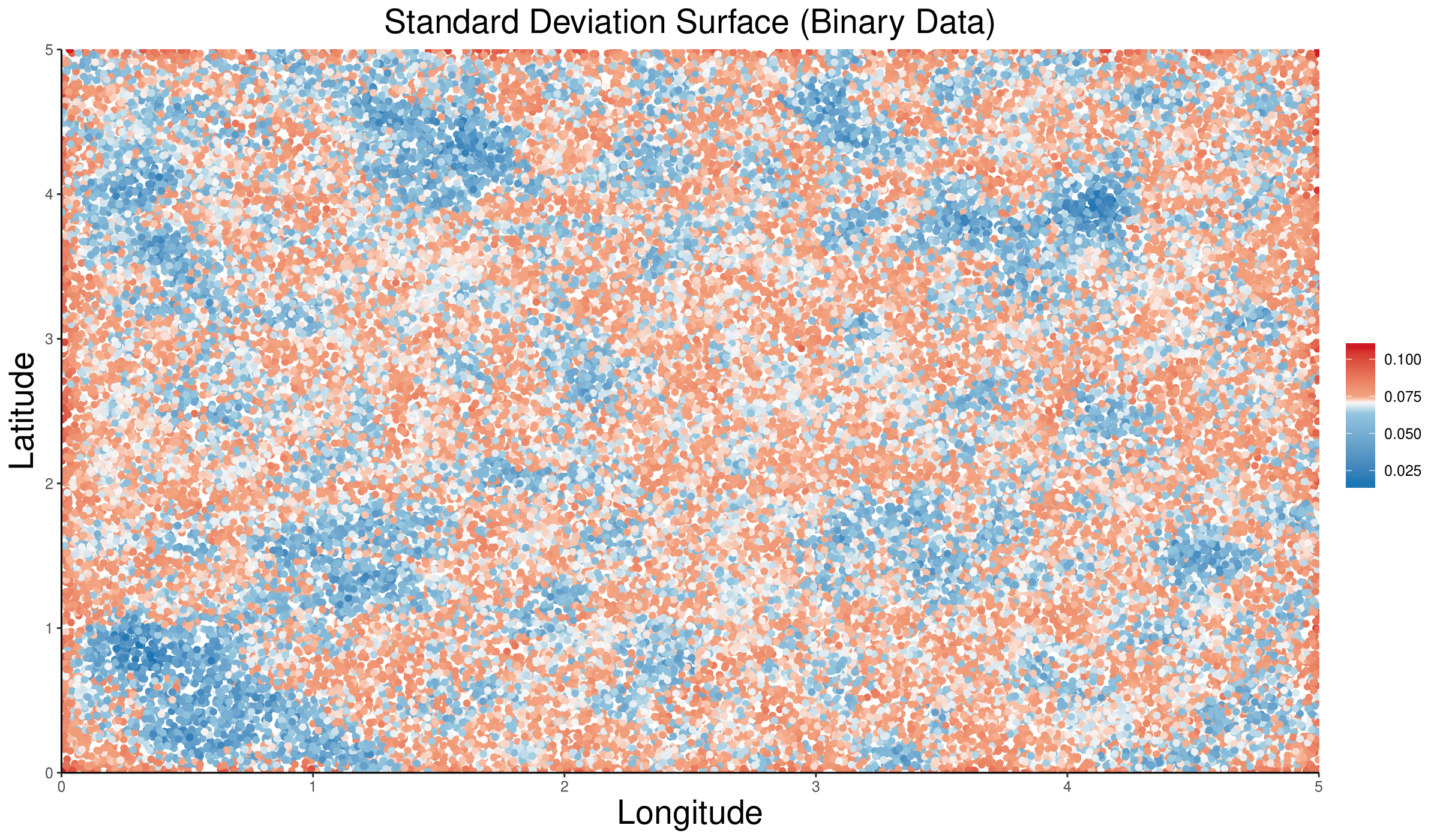}
\caption{Prediction standard deviation surface for the binary example.}
\label{Fig:Binary_SD_Surface}
\end{figure}

\begin{figure}[H]
\centering
\includegraphics[scale=.4]{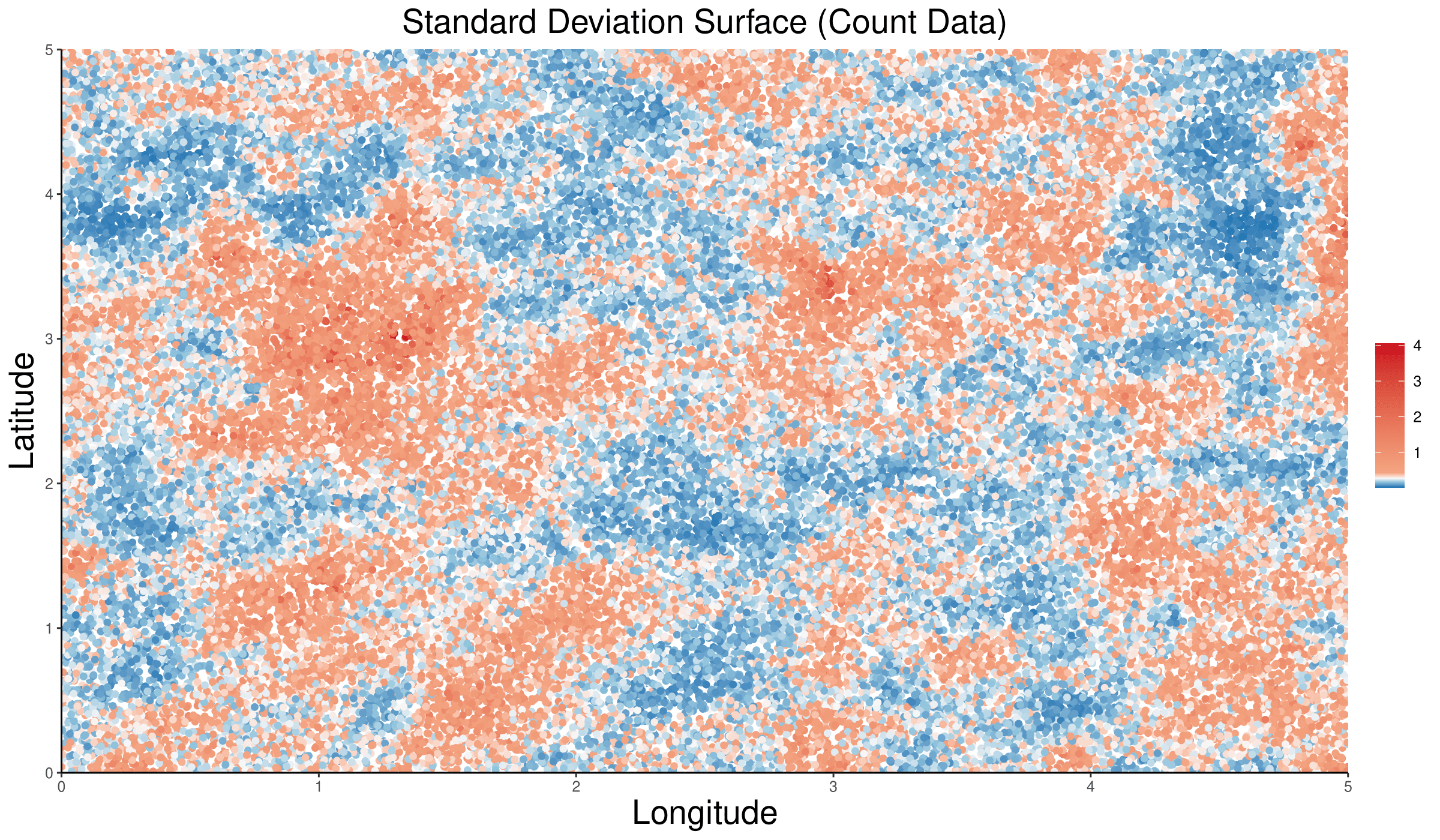}
\caption{Prediction standard deviation surface for the count example.}
\label{Fig:Count_SD_Surface}
\end{figure}

Figure ~\ref{Fig:Binary_SD_Surface} and Figure ~\ref{Fig:Count_SD_Surface} illustrate example prediction standard deviation surfaces for the binary and count examples, respectively. It can be noted that regions with high predicted intensity exhibit correspondingly high standard deviations. Conversely, regions with both high and low predicted probabilities demonstrate lower standard deviations. 

\section{Generation of Nonstationary Spatial Random Effects}
\setcounter{equation}{0}

The nonstationary spatial random effects $\boldsymbol w = \{w(\boldsymbol s_i) :\boldsymbol s_i\in\mathcal D \}$ are generated by smoothing several locally stationary processes contained in disjoint subregions \citep{fuentes2001high}. To do this, we partition the spatial domain $\mathcal D$ into four disjoint subregions $\mathcal D_1$, $\mathcal D_2$, $\mathcal D_3$, and $\mathcal D_4$, where $\mathcal D_1=[0,2.5]^2$, $\mathcal D_2=[2.5,5]^2$, $\mathcal D_3=[0,2.5]\times[2.5,5]$, and $\mathcal D_4=[2.5,5]\times[0,2.5]$. We then specify $C_1(\cdot)$, $C_2(\cdot)$, $C_3(\cdot)$, and $C_4(\cdot)$ to be stationary covariance functions associated with each of the four subregions. Each stationary covariance function comes from the Mat\'ern class with smoothness $\nu=0.5$, partial sill parameter $\sigma^2=1$, and respective range parameters $\phi_1=0.5$, $\phi_2=0.4$, $\phi_3=0.3$, and $\phi_4=0.2$. The nonstationary global covariance function is then constructed using the modeling framework of \cite{nott2002estimation}, where
$$C(\boldsymbol s,\boldsymbol t)=\sum_{i=1}^4 \lambda_i(\boldsymbol s)\lambda_i(\boldsymbol t)C_i(\boldsymbol s,\boldsymbol t).$$
\noindent Here $\lambda_i(\boldsymbol s)$ is a weight function based on the distance between location $\boldsymbol s$ and the center of subregion $\mathcal D_i$ which we denote as $\bm\mu_i$. The weight function is chosen such that $\lambda_i(\boldsymbol s)\geq 0$, $\sum_{i=1}^4\lambda_i(\boldsymbol s)=1$, $\lambda_i(\boldsymbol s)$ attains its maximum at $\bm\mu_i$, and decays smoothly to zero as $||\boldsymbol s - \bm\mu_i||\rightarrow\infty $. To ensure that $\lambda_i(\boldsymbol s)\geq 0$, \cite{nott2002estimation} employ the kernel function
$$\kappa_{\eta}(\boldsymbol t)=\exp\left(-\frac{||\boldsymbol t||^2}{\eta}\right),$$

\noindent where $\eta$ is a smoothing parameter (we specify $\eta=6$), and then
$$\lambda_i(\boldsymbol s)=\frac{\kappa_{\eta}(\boldsymbol s-\bm\mu_i)}{\sum_{j=1}^4 \kappa_{\eta}(\boldsymbol s-\bm\mu_j)}.$$

\section{Posterior Difference Surfaces}

For one simulated nonstationary dataset, Figure ~\ref{Fig:Post_Diff_Bin_50K} and Figure ~\ref{Fig:Post_Diff_Pois_50K} below display the posterior difference in the log intensities for the count data case, as well as the posterior difference in probabilities for the binary data case. There does not appear to be any discernible spatial pattern in the difference surfaces in either case. These are due to properties of the binomial and Poisson distribution.

\begin{figure}[H]
\centering
\includegraphics[scale=.4]{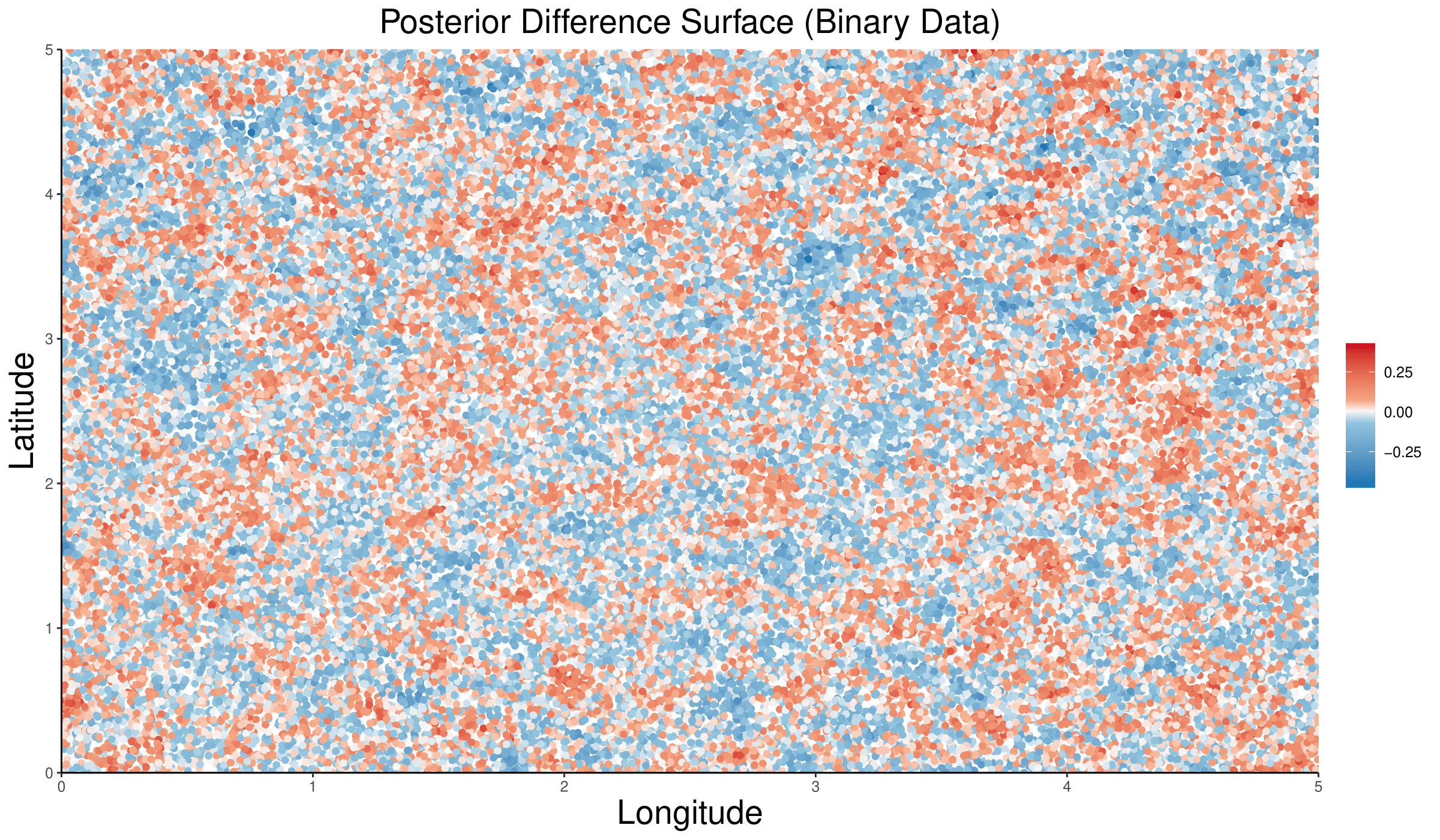}
\caption{Posterior difference of the probability.}
\label{Fig:Post_Diff_Bin_50K}
\end{figure} 

\begin{figure}[H]
\centering
\includegraphics[scale=.4]{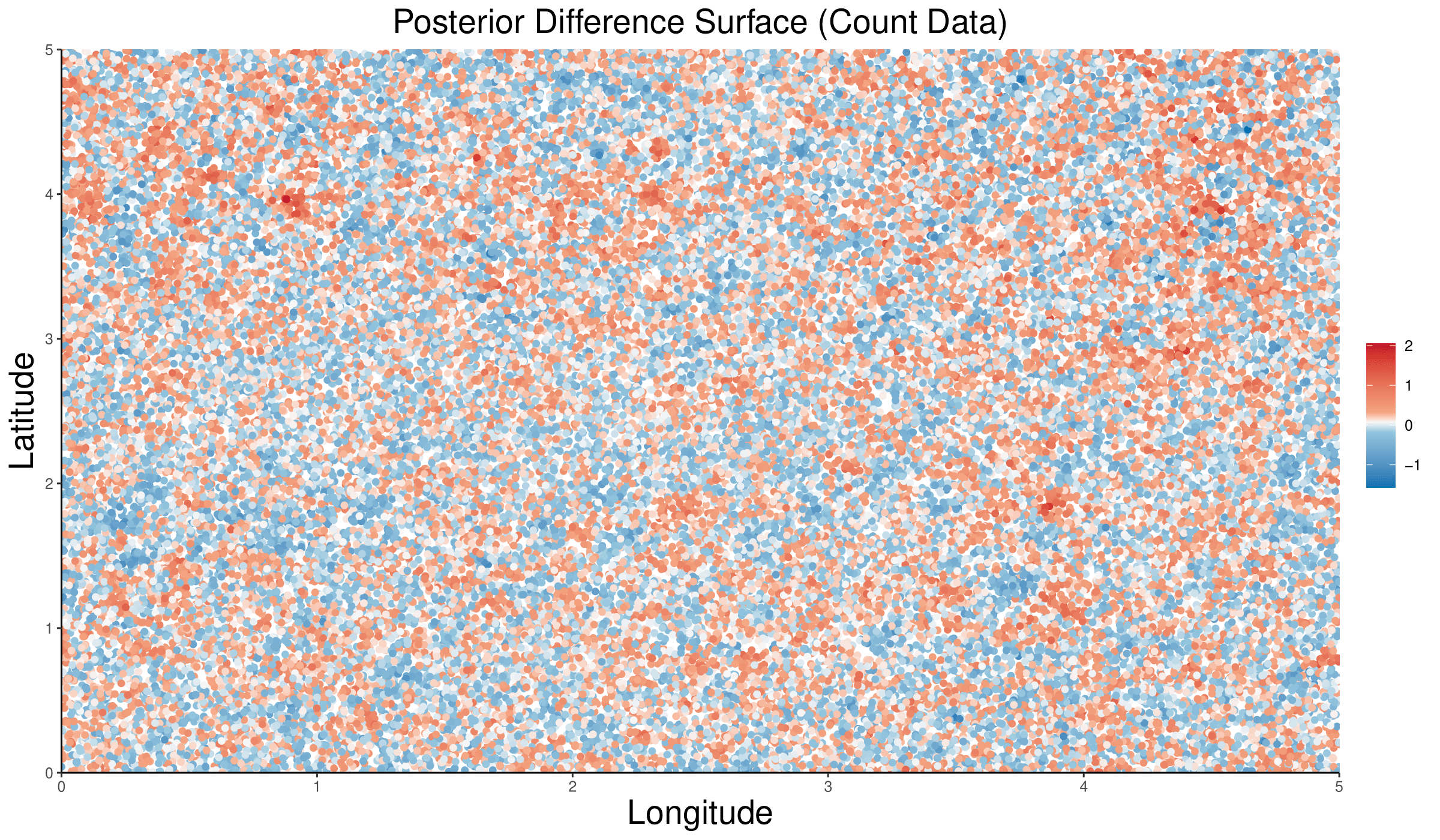}
\caption{Posterior difference of the log intensity.}
\label{Fig:Post_Diff_Pois_50K}
\end{figure} 

\section{Coverage Surfaces}

\begin{figure}[H]
\centering
\includegraphics[scale=.4]{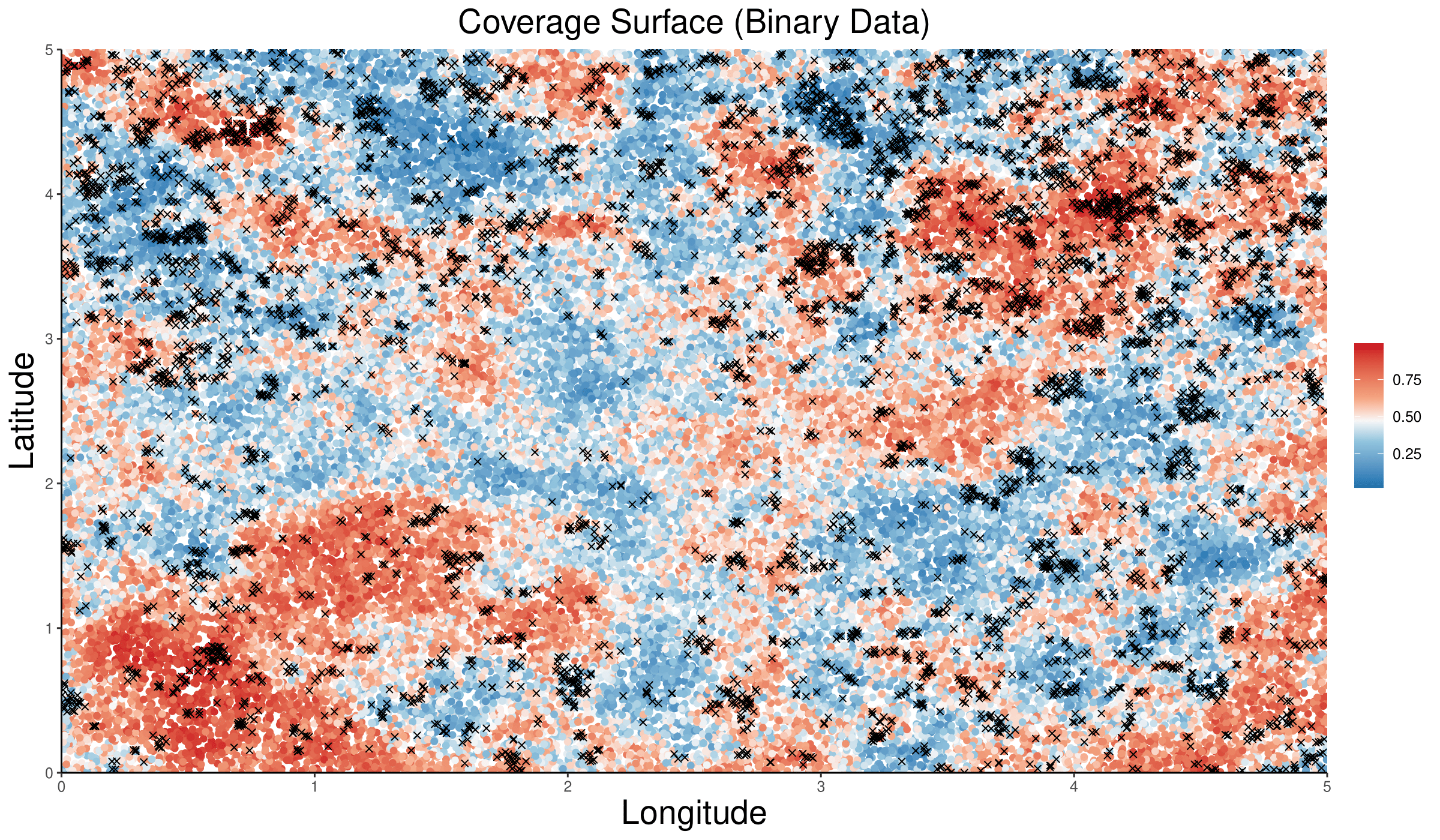}
\caption{Coverage of Credible Intervals. Each `x' represents a location where the 95\% credible interval does not include the true probability.}
\label{Fig:Covered_Bin_50K}
\end{figure}

\begin{figure}[H]
\centering
\includegraphics[scale=.4]{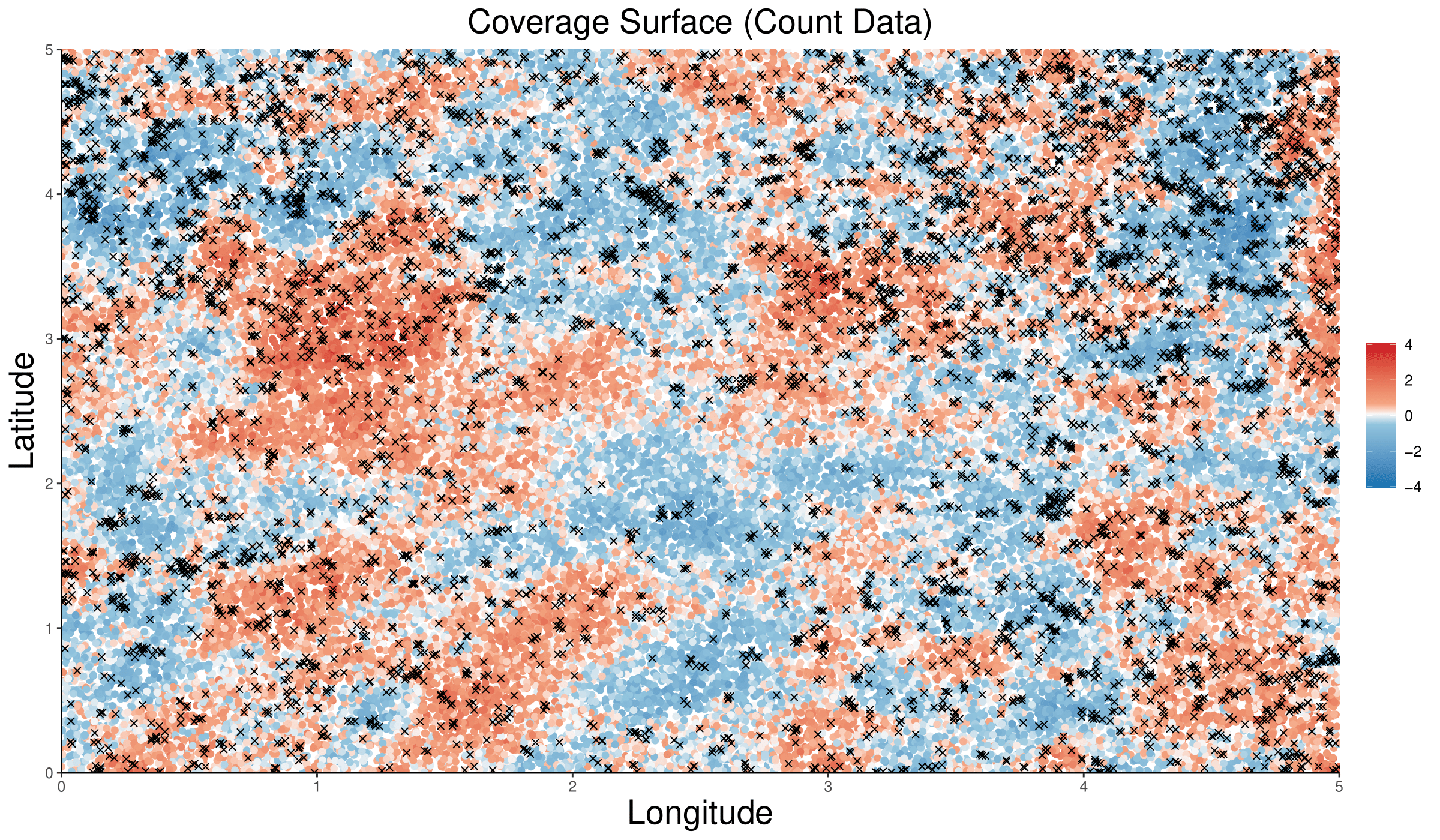}
\caption{Coverage of Credible Intervals. Each `x' represents a location where the 95\% credible interval does not include the true log intensity.}
\label{Fig:Covered_Pois_50K}
\end{figure}

Figure ~\ref{Fig:Covered_Bin_50K} below indicates locations where the true probability is not covered by the 95\% credible interval. Similarly, Figure ~\ref{Fig:Covered_Pois_50K} indicates locations where the true log intensity is not covered by the 95\% credible interval.
The respective coverage probabilities are 89.58\% and 89.22\%, for the binary and count datasets respectively. These are similar to the coverage probabilities from the bisquare basis approach, 88.25\% 87.78\%. 

It is worth noting that regions with higher and lower probabilities in the binary case have smaller coverage probabilities. This is likely due to the lower standard deviations (smaller coverage intervals) associated with higher and lower probabilities. On the other hand, regions with higher log intensities exhibit smaller coverage probabilities, which is interesting. This could potentially be due to the right skewness of the Poisson distribution.

\section{Procedure for choosing $K$}

For practitioners, we suggest searching over a range of possible values. Specifically $K\in\{K_{\text{min}},\ldots,K_{\text{max}}\}$, where $K_{\text{min}}=\max\{4,\lceil\log_{10}(n)\rceil\}$, and $K_{\text{max}}$ is the largest value of $K$ such that each partition contains at least 50 observations. We set the minimum value of $K$ to be at least four because there are four global basis functions with the largest bandwidths. If cross validation is not feasible, one could use Akaike's information criterion (AIC), the Schwarz-Bayesian criterion (BIC), or adjusted $R^2$ from a model that includes the covariates, global basis functions, and cluster assignment as predictors. For instance, one can initially cluster observations using $K=\max\{4,\lceil\log_{10}(n)\rceil\}$ clusters and then incrementally increase $K$ either until the decrease in AIC is negligible (using the so-called ``elbow" method) or until $K_{\text{max}}$ is reached. An example of this procedure is provided in the supplement. In our simulation study, we compare the performance of our method with various choices of $K$. In fact, our method is fast enough such that practitioners can explore multiple $K$ settings and ultimately choose the most accurate $K$ based on out-of-sample predictions.

\begin{figure}[H]
\centering
\includegraphics[scale=.4]{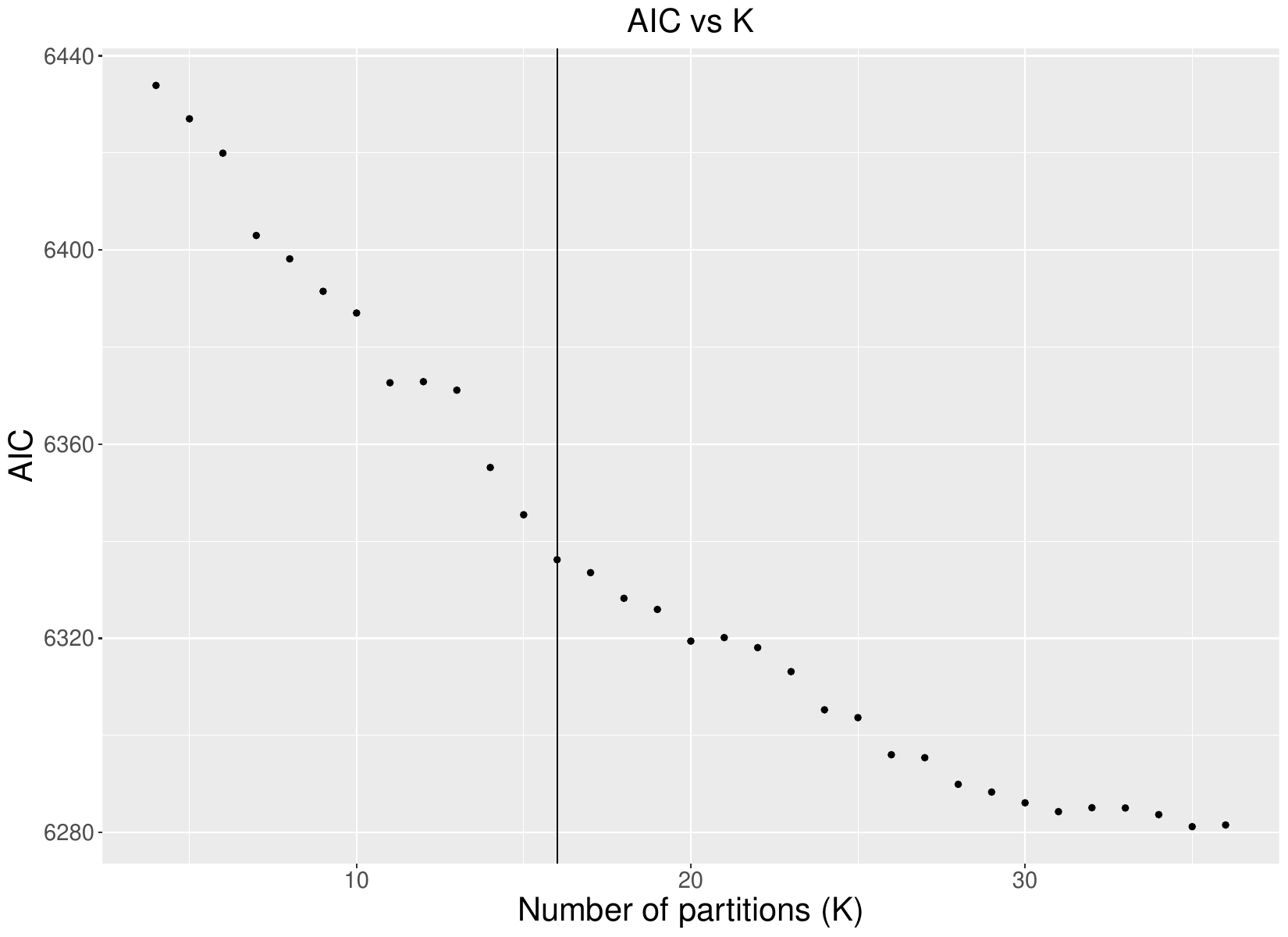}
\caption{Elbow method for providing an initial estimate for the number of partitions $K$.}
\label{Fig:elbow}
\end{figure}

We have implemented an example of this procedure on one nonstationary spatial dataset with 5{,}000 observations. As displayed in Figure \ref{Fig:elbow}, an initial guess for the number of partitions $K$ would be 16 based on the elbow method. 

\section{Standard Deviation Surfaces for the Real-World Applications}
\setcounter{equation}{0}

\begin{figure}[H]
\centering
\includegraphics[scale=.4]{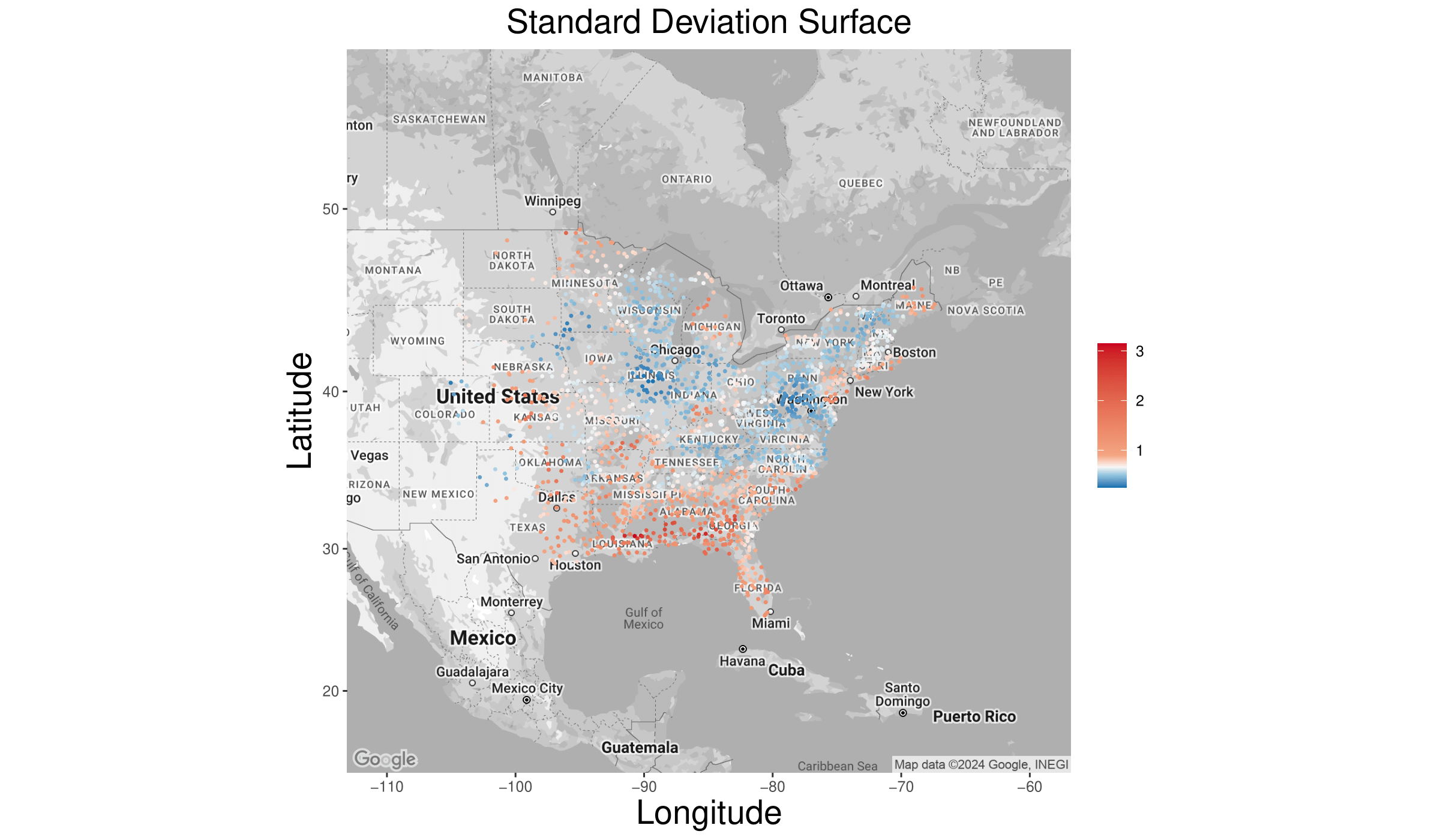}
\caption{Prediction standard deviation surface for the Blue Jay count data example.}
\label{Fig:BlueJay_SD_Surface}
\end{figure}

\begin{figure}[H]
\centering
\includegraphics[scale=.4]{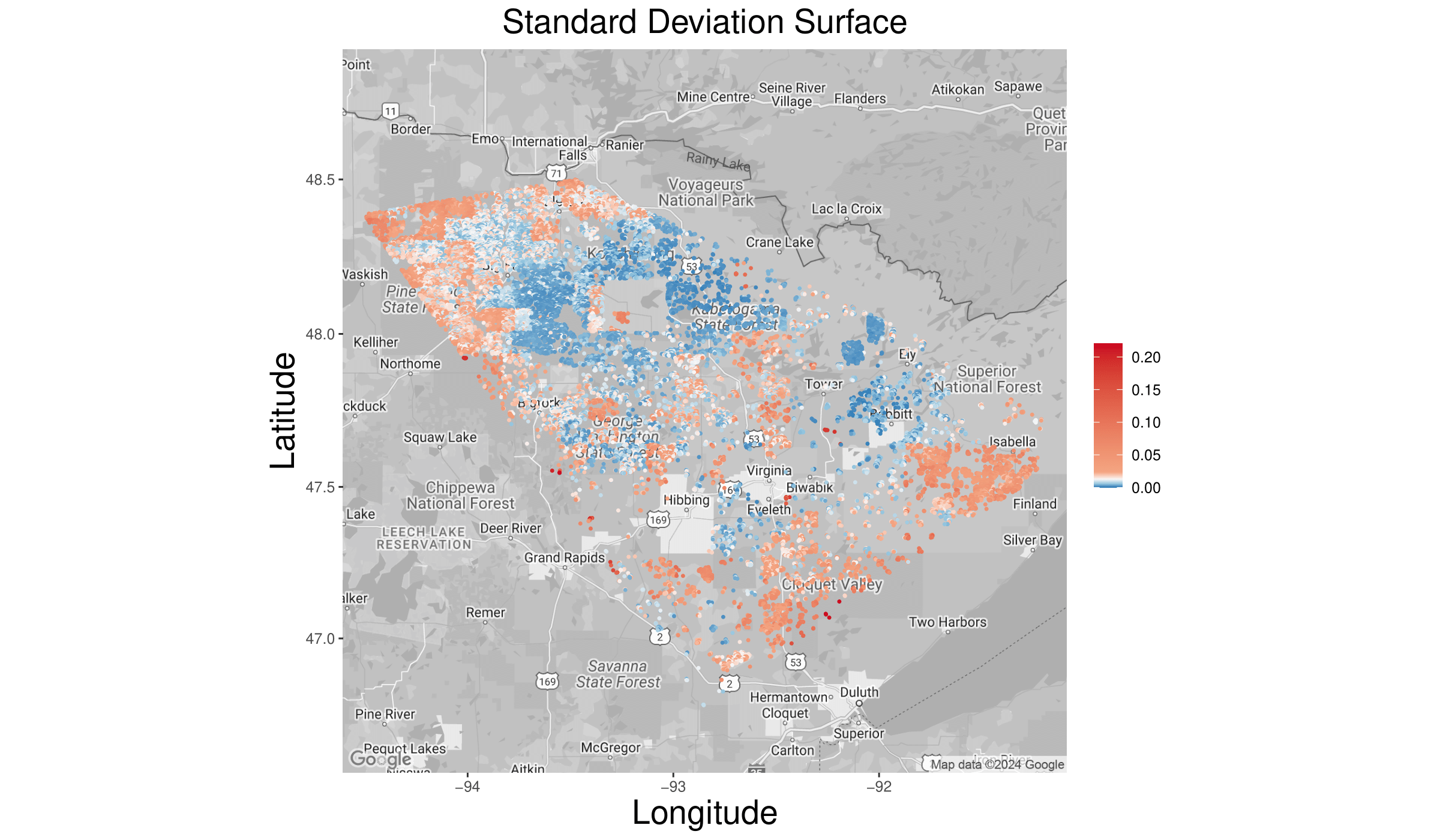}
\caption{Prediction standard deviation surface for the dwarf mistletoe binary data example.}
\label{Fig:Mistle_SD_RdBu}
\end{figure}

Figure ~\ref{Fig:BlueJay_SD_Surface} and Figure ~\ref{Fig:Mistle_SD_RdBu} illustrate the prediction standard deviation surfaces for the Blue Jay abundance and dwarf mistletoe incidence applications, respectively. It can be noted that regions with high predicted intensity exhibit correspondingly high standard deviations for the Blue Jay example. Conversely, regions with both high and low predicted probabilities demonstrate lower standard deviations for the dwarf mistletoe example. These are due to properties of the binomial and Poisson distribution.

\section{Inference results for the Dwarf Mistletoe Example}
\setcounter{equation}{0}

Table ~\ref{Tab2:Dwarf_Cov} below displays the parameter estimates for the regression coefficients in the dwarf mistletoe example. Model-fitting results suggest that tree stands with older (higher age) and taller (higher canopy height) dominant trees have a greater probability of dwarf mistletoe infestation. Conversely, tree stands with a higher basal area per acre and larger stand volume have a lower probability of dwarf mistletoe presence. Each variable is found to be significant, except for basal area, for which the 95\% credible interval contains 0. 

\begin{table}[H]
\centering
\begin{tabular}{>{\RaggedLeft}p{2cm}>{\RaggedLeft}p{2cm}p{2.5cm}}
\toprule
Covariate & Estimate & 95\% CI \\
\midrule
Age & $1.10$ & (0.76,1.43) \\
Basal Area & $-0.41$ & (-1.57,0.60)\\
Height & 2.70 & (2.42,2.99)\\
Volume & -1.15 & (-1.66,-0.61)\\
\bottomrule
\end{tabular}
\caption{Inference results for the mistletoe data. Rows correspond to the predictor variables and columns include the parameter estimates and 95\% credible intervals.}
\label{Tab2:Dwarf_Cov}
\end{table}